\documentclass[twocolumn]{aastex701}
\usepackage{amsmath}

\newcommand{\sersic}{S\'{e}rsic\,}

\begin{document}


\title{

VENUS:\\
Two Faint Little Red Dots Separated by $\sim{150}\,\mathrm{pc}$ \\
Hidden in a Single Lensed Galaxy at $z\sim7$

%
%
}


\author[orcid=0009-0006-6763-4245]{Hiroto Yanagisawa}
\affiliation{Institute for Cosmic Ray Research, The University of Tokyo, 5-1-5 Kashiwanoha, Kashiwa, Chiba 277-8582, Japan}
\affiliation{Department of Physics, Graduate School of Science, The University of Tokyo, 7-3-1 Hongo, Bunkyo, Tokyo 113-0033, Japan}
\email[show]{yana@icrr.u-tokyo.ac.jp}

\author[0000-0002-1049-6658]{Masami Ouchi}
\affiliation{National Astronomical Observatory of Japan, 2-21-1 Osawa, Mitaka, Tokyo 181-8588, Japan}
\affiliation{Institute for Cosmic Ray Research, The University of Tokyo, 5-1-5 Kashiwanoha, Kashiwa, Chiba 277-8582, Japan}
\affiliation{Department of Astronomical Science, SOKENDAI (The Graduate University for Advanced Studies), 2-21-1 Osawa, Mitaka, Tokyo, 181-8588, Japan}
\affiliation{Kavli Institute for the Physics and Mathematics of the Universe (WPI), The University of Tokyo, 5-1-5 Kashiwanoha, Kashiwa, Chiba 277-8583, Japan}
\email[]{ouchims@icrr.u-tokyo.ac.jp}

\author[0000-0001-9411-3484]{Miriam Golubchik}
\affiliation{Department of Physics, Ben-Gurion University of the Negev, P.O. Box 653, Beer-Sheva 8410501, Israel}
\email[]{golubmir@post.bgu.ac.il}

\author[0000-0003-3484-399X]{Masamune Oguri}
\affiliation{Center for Frontier Science, Chiba University, 1-33 Yayoi-cho, Inage-ku, Chiba 263-8522, Japan}
\affiliation{Department of Physics, Graduate School of Science, Chiba University, 1-33 Yayoi-Cho, Inage-Ku, Chiba 263-8522, Japan}
\email[]{masamune.oguri@chiba-u.jp}

\author[0000-0001-7201-5066]{Seiji Fujimoto}
\affiliation{David A. Dunlap Department of Astronomy and Astrophysics, University of Toronto, 50 St. George Street, Toronto, Ontario, M5S 3H4, Canada}
\affiliation{Dunlap Institute for Astronomy and Astrophysics, 50 St. George Street, Toronto, Ontario, M5S 3H4, Canada}
\email[]{seiji.fujimoto@utoronto.ca}

\author[0000-0002-5588-9156]{Vasily Kokorev}
\affiliation{Department of Astronomy, The University of Texas at Austin, 2515 Speedway Blvd Stop C1400, Austin, TX 78712, USA}
\affiliation{Cosmic Frontier Center, The University of Texas at Austin, Austin, TX 78712, USA}
\email[]{vkokorev@utexas.edu}

\author[0000-0003-2680-005X]{Gabriel Brammer}
\affiliation{Cosmic Dawn Center (DAWN), Jagtvej 128, DK2200 Copenhagen N, Denmark}
\affiliation{Niels Bohr Institute, University of Copenhagen, Jagtvej 128, 2200 Copenhagen N, Denmark}
\email[]{gabriel.brammer@nbi.ku.dk}

\author[0000-0002-4622-6617]{Fengwu Sun}
\affiliation{Center for Astrophysics $|$ Harvard \& Smithsonian, 60 Garden St., Cambridge, MA 02138, USA}
\email[]{fengwu.sun@cfa.harvard.edu}

\author[0009-0000-1999-5472]{Minami Nakane}
\affiliation{Institute for Cosmic Ray Research, The University of Tokyo, 5-1-5 Kashiwanoha, Kashiwa, Chiba 277-8582, Japan}
\affiliation{Department of Physics, Graduate School of Science, The University of Tokyo, 7-3-1 Hongo, Bunkyo, Tokyo 113-0033, Japan}
\email[]{nakanem@icrr.u-tokyo.ac.jp}

\author[0000-0002-6047-430X]{Yuichi Harikane}
\affiliation{Institute for Cosmic Ray Research, The University of Tokyo, 5-1-5 Kashiwanoha, Kashiwa, Chiba 277-8582, Japan}
\email[]{hari@icrr.u-tokyo.ac.jp}

\author[0009-0008-0167-5129]{Hiroya Umeda}
\affiliation{Institute for Cosmic Ray Research, The University of Tokyo, 5-1-5 Kashiwanoha, Kashiwa, Chiba 277-8582, Japan}
\affiliation{Department of Physics, Graduate School of Science, The University of Tokyo, 7-3-1 Hongo, Bunkyo, Tokyo 113-0033, Japan}
\email[]{ume@icrr.u-tokyo.ac.jp}

\author[0000-0003-3596-8794]{Hollis B. Akins}
\affiliation{Department of Astronomy, The University of Texas at Austin, 2515 Speedway Blvd Stop C1400, Austin, TX 78712, USA}

\email[]{hollis.akins@gmail.com}

\author[0000-0002-7570-0824]{Hakim Atek}
\affiliation{Institut d'Astrophysique de Paris, CNRS, Sorbonne Universit\'e, 98bis Boulevard Arago, 75014, Paris, France}
\email[]{hakim.atek@iap.fr}

\author[0000-0002-8686-8737]{Franz E. Bauer}
\affiliation{Instituto de Alta Investigaci{\'{o}}n, Universidad de Tarapac{\'{a}}, Casilla 7D, Arica, 1010000, Chile}
\email[]{fbauer@academicos.uta.cl}

\author[0000-0001-5984-0395]{Maru{\v{s}}a Brada{\v{c}}}
\affiliation{University of Ljubljana, Faculty of Mathematics and Physics, Jadranska ulica 19, SI-1000 Ljubljana, Slovenia}
\affiliation{Department of Physics and Astronomy, University of California Davis, 1 Shields Avenue, Davis, CA 95616, USA}
\email[]{marusa.bradac@fmf.uni-lj.si}

\author[0000-0002-0302-2577]{John Chisholm}
\affiliation{Department of Astronomy, The University of Texas at Austin, 2515 Speedway Blvd Stop C1400, Austin, TX 78712, USA}
\affiliation{Cosmic Frontier Center, The University of Texas at Austin, Austin, TX 78712, USA}
\email[]{chisholm@austin.utexas.edu}

\author[0000-0001-7410-7669]{Dan Coe}
\affiliation{Space Telescope Science Institute, 3700 San Martin Drive, Baltimore, MD 21218, USA}
\affiliation{Association of Universities for Research in Astronomy (AURA), Inc.~for the European Space Agency (ESA)}
\affiliation{Center for Astrophysical Sciences, Department of Physics and Astronomy, The Johns Hopkins University, 3400 N Charles St. Baltimore, MD 21218, USA}
\email[]{dcoe@stsci.edu}

\author[0000-0001-9065-3926]{Jose M. Diego}
\affiliation{Instituto e F\'isica de Cantabria,(CSIC-UC), Avda. Los Castros s/n. 39005, Santander, Spain}
\email[]{jdiego@ifca.unican.es}

\author[0000-0001-7113-2738]{Henry C. Ferguson}
\affiliation{Space Telescope Science Institute, 3700 San Martin Drive, Baltimore, MD 21218, USA}
\email[]{ferguson@stsci.edu}

\author[0000-0001-8519-1130]{Steven L. Finkelstein}
\affiliation{Department of Astronomy, The University of Texas at Austin, 2515 Speedway Blvd Stop C1400, Austin, TX 78712, USA}
\affiliation{Cosmic Frontier Center, The University of Texas at Austin, Austin, TX 78712, USA}
\email[]{stevenf@astro.as.utexas.edu}

\author[orcid=0000-0001-6278-032X]{Lukas J. Furtak}
\affiliation{Department of Astronomy, The University of Texas at Austin, 2515 Speedway Blvd Stop C1400, Austin, TX 78712, USA}
\affiliation{Cosmic Frontier Center, The University of Texas at Austin, Austin, TX 78712, USA}
\email[]{furtak@utexas.edu}

\author[orcid=0000-0001-9840-4959,sname='']{Kohei Inayoshi} \affiliation{Kavli Institute for Astronomy and Astrophysics, Peking University, Beijing 100871, China}
\email[]{inayoshi@pku.edu.cn}

\author[0000-0003-2871-127X]{Jorryt Matthee}  \affiliation{Institute of Science and Technology Austria (ISTA), Am Campus 1, 3400 Klosterneuburg, Austria}
\email[]{jorryt.matthee@ista.ac.at}

\author[0000-0003-3729-1684]{Rohan P. Naidu}
\affiliation{MIT Kavli Institute for Astrophysics and Space Research, 70 Vassar Street, Cambridge, MA 02139, USA}
\email[]{rnaidu@mit.edu}

\author[0000-0001-9011-7605]{Yoshiaki Ono}
\affiliation{Institute for Cosmic Ray Research, The University of Tokyo, 5-1-5 Kashiwanoha, Kashiwa, Chiba 277-8582, Japan}
\email[]{ono@icrr.u-tokyo.ac.jp}

\author[0000-0002-9651-5716]{Richard Pan}
\affiliation{Department of Physics and Astronomy, Tufts University, 574 Boston Avenue, Suite 304, Medford, MA 02155, USA}
\email[]{richard.pan@tufts.edu}

\author[0000-0001-5492-1049]{Johan Richard}
\affiliation{Univ Lyon, Univ Lyon1, ENS de Lyon, CNRS, Centre de Recherche Astrophysique de Lyon UMR5574, 69230 Saint-Genis-Laval, France}
\email[]{johan.richard@univ-lyon1.fr}

\author[0000-0002-6265-2675]{Luke Robbins}
\affiliation{Department of Physics and Astronomy, Tufts University, 574 Boston Avenue, Suite 304, Medford, MA 02155, USA}
\email[]{andrew.robbins@tufts.edu}

\author[0000-0002-4201-7367]{Chris Willott} \affiliation{NRC Herzberg, 5071 West Saanich Rd, Victoria, BC V9E 2E7, Canada}
\email[]{chris.willott@nrc.ca}

\author[orcid=0000-0002-0350-4488]{Adi Zitrin}
\affiliation{Department of Physics, Ben-Gurion University of the Negev, P.O. Box 653, Beer-Sheva 8410501, Israel}
\email[]{zitrin@bgu.ac.il}

\author[0000-0001-5758-1000]{Ricardo O. Amor\'{i}n} 
\affiliation{Instituto de Astrof\'{i}sica de Andaluc\'{i}a (CSIC), Apartado 3004, 18080 Granada, Spain}
\email[]{amorin@iaa.es}

\author[0000-0002-7908-9284]{Larry D. Bradley}
\affiliation{Space Telescope Science Institute, 3700 San Martin Drive, Baltimore, MD 21218, USA}
\email[]{larry.bradley@gmail.com}

\author[0000-0003-0212-2979]{Volker Bromm}
\affiliation{Department of Astronomy, The University of Texas at Austin, 2515 Speedway Blvd Stop C1400, Austin, TX 78712, USA}
\affiliation{Cosmic Frontier Center, The University of Texas at Austin, Austin, TX 78712, USA}
\affiliation{Weinberg Institute for Theoretical Physics, University of Texas, Austin, TX 78712, USA}
\email[]{vbromm@astro.as.utexas.edu}

\author[0000-0003-1949-7638]{Christopher J. Conselice}
\affiliation{Jodrell Bank Centre for Astrophysics, University of Manchester, Oxford Road, Manchester M13 9PL, UK}
\email[]{conselice@manchester.ac.uk}

\author[0000-0001-8460-1564]{Pratika Dayal}
\affiliation{Canadian Institute for Theoretical Astrophysics, 60 St George St, University of Toronto, Toronto, ON M5S 3H8, Canada}
\affiliation{David A. Dunlap Department of Astronomy and Astrophysics, University of Toronto, 50 St. George Street, Toronto, Ontario, M5S 3H4, Canada}
\affiliation{Department of Physics, 60 St George St, University of Toronto, Toronto, ON M5S 3H8, Canada}
\email[]{pratika.dayal@utoronto.ca}

\author[0000-0001-7232-5355]{Qinyue Fei} \affiliation{David A. Dunlap Department of Astronomy and Astrophysics, University of Toronto, 50 St. George Street, Toronto, Ontario, M5S 3H4, Canada}
\email[]{qy.fei@utoronto.ca}

\author[0000-0002-9382-9832]{Andreas L. Faisst}
\email{afaisst@caltech.edu}
\affiliation{IPAC, California Institute of Technology, 1200 E. California Blvd. Pasadena, CA 91125, USA}

\author[0000-0001-9187-3605]{Jeyhan S. Kartaltepe}
\affiliation{Laboratory for Multiwavelength Astrophysics, School of Physics and Astronomy, Rochester Institute of Technology, 84 Lomb Memorial Drive, Rochester, NY 14623, USA}
\email[]{jeyhan@astro.rit.edu}

\author[0000-0003-2540-7424]{Paulo A. A. Lopes}
\affiliation{Observatório do Valongo, Universidade Federal do Rio de Janeiro, Ladeira do Pedro Antônio 43, Rio de Janeiro, RJ, 20080-090, Brazil}
\email[]{plopes@ov.ufrj.br}

\author[0000-0003-1581-7825]{Ray A. Lucas}
\affiliation{Space Telescope Science Institute, 3700 San Martin Drive, Baltimore, MD 21218, USA}
\email[]{lucas@stsci.edu}

\author[0000-0002-4872-2294]{Georgios E. Magdis}
\affiliation{Cosmic Dawn Center (DAWN), Jagtvej 128, DK2200 Copenhagen N, Denmark}
\affiliation{DTU-Space, Technical University of Denmark, Elektrovej 327, 2800, Kgs. Lyngby, Denmark}
\email[]{geoma@space.dtu.dk}

\author[orcid=0000-0003-3243-9969]{Nicholas S. Martis}
\affiliation{University of Ljubljana, Faculty of Mathematics and Physics, Jadranska ulica 19, SI-1000 Ljubljana, Slovenia}
\email[]{nicholas.martis@fmf.uni-lj.si}

\author[0000-0001-7503-8482]{Casey Papovich}
\affiliation{Department of Physics and Astronomy, Texas A\&M University, College Station, TX 77843-4242, USA}
\affiliation{George P. and Cynthia Woods Mitchell Institute for Fundamental Physics and Astronomy, Texas A\&M University, College Station, TX 77843-4242, USA}
\email{papovich@tamu.edu}

\author[0000-0001-7144-7182]{Daniel Schaerer}
\affiliation{Department of Astronomy, University of Geneva, Chemin Pegasi 51, 1290 Versoix, Switzerland}
\email[]{daniel.schaerer@unige.ch}

\author[0000-0001-6477-4011]{Francesco Valentino}
\affiliation{Cosmic Dawn Center (DAWN), Jagtvej 128, DK2200 Copenhagen N, Denmark}
\affiliation{DTU-Space, Technical University of Denmark, Elektrovej 327, 2800, Kgs. Lyngby, Denmark}
\email[]{fmava@dtu.dk}

\author[0000-0002-5057-135X]{Eros Vanzella}
\affiliation{INAF -- OAS, Osservatorio di Astrofisica e Scienza dello Spazio di Bologna, via Gobetti 93/3, I-40129 Bologna, Italy}
\email[]{eros.vanzella@inaf.it}

\author[0000-0003-2718-8640]{Joseph F. V. Allingham}
\affiliation{Department of Physics, Ben-Gurion University of the Negev, P.O. Box 653, Beer-Sheva 8410501, Israel}
\email[]{allingha@post.bgu.ac.il}

\author[0000-0001-9440-8872]{Norman A. Grogin}
\affiliation{Space Telescope Science Institute, 3700 San Martin Drive, Baltimore, MD 21218, USA}
\email[]{nagrogin@stsci.edu}

\author[0000-0002-4837-1615]{Mauro González-Otero}
\affiliation{Instituto de Astrof\'isica de Andaluc\'ia--CSIC, Glorieta de la Astronom\'ia s/n, E--18008 Granada, Spain}
\email[]{mauromarago@gmail.com}

\author[0000-0003-4223-7324]{Massimo Ricotti}
\affiliation{Department of Astronomy, University of Maryland, College Park, 20742, USA}
\email[]{ricotti@umd.edu}

\author[0000-0001-8156-6281]{Rogier A. Windhorst}
\affiliation{School of Earth and Space Exploration, Arizona State University,
Tempe, AZ 85287-6004, USA} \email[]{Rogier.Windhorst@asu.edu}

\begin{abstract}
We report the identification of a pair of faint little red dots (LRDs), dubbed Red Eyes, in a strongly-lensed galaxy at $z\sim7$ behind the PLCK {G004.5-19.5} cluster, identified from the JWST Treasury program VENUS. Red Eyes are spatially resolved on the image plane with distinct colors, while the critical curve lies far north of Red Eyes, clearly requiring two different LRDs rather than a single LRD. Red Eyes is an extremely close pair of LRDs separated by $\sim {150}\,\mathrm{pc}$ in the source plane with a magnification of $\mu\sim{11}$, which consistently explains another counter-image detected to the north-west. Red Eyes is hosted in a typical star-forming galaxy with $M_{\mathrm{UV,int}}\sim -19$, but its own UV emission is very faint ($M_{\mathrm{UV,int}} \gtrsim {-17}$). Moreover, Red Eyes lies at an offset position from the galaxy component. If observed without lensing, Red Eyes would appear as a typical star-forming galaxy at $z\sim 7$ with $M_{\mathrm{UV}}\sim -19$,
{without unambiguous LRD signatures in either morphology or SED.} These results suggest that multiple off-center LRDs, similar to Red Eyes, may be commonly hidden in a typical high-$z$ star-forming galaxy. In this case, various plausible scenarios may emerge, one of which is that intermediate-mass black holes (IMBHs) with $M_\mathrm{BH}\sim10^{4\text{--}6}\,M_\odot$ may form in star clusters on a stellar disk and contribute to the growth of the central supermassive black hole via mergers, with some IMBHs detectable as luminous LRDs in a sufficiently active and massive phase.

\end{abstract}

\keywords{}

\section{Introduction}
James Webb Space Telescope (JWST) has identified a population of broad-line active galactic nuclei (AGNs) that exhibit a characteristic V-shaped spectral energy distribution (SED), with a blue UV continuum and a red optical continuum (e.g., \citealp{Kocevski+2023, Harikane+2023, Matthee+2024, Kocevski+2025, Labbe+2025, Kokorev+2024, Akins+2025, Greene+2024, Williams+2024, Tanaka+2025, Inayoshi_Ho_2025}). These objects, commonly referred to as little red dots (LRDs), show extremely weak X-ray, mid-infrared, and radio emission, which challenges conventional AGN interpretations \citep{Yue+2024, Akins+2025, Ananna+2024, Gloudemans+2025, Maiolino+2025}. While many LRDs exhibit only weak variability \citep{Burke+2025, Kokubo_Harikane_2025, Stone+2025, Tee+2025, ZZhang+2025a}, a multiply lensed LRD with a long time delay has been reported to show significant variability \citep{ZZhang+2025b}. First identified at $z\gtrsim4$, LRDs have recently also been found in the local universe, although with a significantly lower number density \citep{Izotov_Thuan_2008, XLin+2025, RLin+2025, XChen+2025}.

The several examples of strong Balmer breaks and Balmer absorption in the LRD spectra (e.g., \citealp{Naidu+2025, deGraaff+2025a, Ji+2025, Ronayne+2025}) have motivated the hypothesis that extremely dense neutral gas surrounding these candidate AGNs can produce such strong Balmer absorption features (e.g., \citealp{Inayoshi_Maiolino_2025, Kido+2025, Inayoshi+2025}). Theoretical studies suggest that these systems may represent an early accretion phase of AGNs \citep{Inayoshi+2025a}, making them a potentially crucial population for understanding the growth of supermassive black holes (SMBHs) and their coevolution with host galaxies. 

These theoretical hypotheses assume that the red optical continuum arises from blackbody emission produced by the dense gas envelope surrounding the AGN, whereas the blue UV continuum originates from young stars in the associated galaxy. Although several studies have performed morphological decompositions of LRDs and their associated galaxy \citep{Killi+2024, Rinaldi+2025, Merida+2025, Zhang+2026}, isolating the galaxy components and deriving their physical properties remain challenging because of the extremely compact nature of LRDs. Several studies have also identified LRDs with blue companions with a few hundred pc to kpc scale offsets \citep{Merida+2025, Golubchik+2025, Baggen+2025, CChen+2025a, CChen+2025b}, which might be related to the UV emission from the LRDs. Gravitationally lensed LRDs are powerful probe for investigating the faint emission from the LRDs \citep{Furtak+2023, Furtak+2024, Merida+2025, Golubchik+2025, Baggen+2025}. However, the coevolution of the LRDs and the associated galaxy, or their connection to the SMBH growth, are still unclear. 



In this work, we report the discovery of a pair of strongly lensed LRDs, dubbed Red Eyes, and an associated galaxy identified at $z\sim7$ in PLCK {G004.5-19.5} cluster by JWST/NIRCam observations. In Section \ref{sec:obs}, we present our observations and data reduction. In Section \ref{sec:analysis}, we describe our analysis methods. Section \ref{sec:results} presents the results and their implications. We present our conclusions in Section \ref{sec:summary}. Throughout this paper, we assume cosmology parameters based on the TT, TE, EE + lowE + lensing + BAO result from \cite{Planck+2020} with $H_0=67.66 \, \mathrm{km\,s^{-1}\,Mpc^{-1}}$, $\Omega_\mathrm{m}=0.30966$, and $\Omega_\mathrm{b}=0.04897$. All magnitudes are in the AB system \citep{Oke_Gunn_1983}.

\begin{figure*}
    \centering
    \includegraphics[width=1\linewidth]{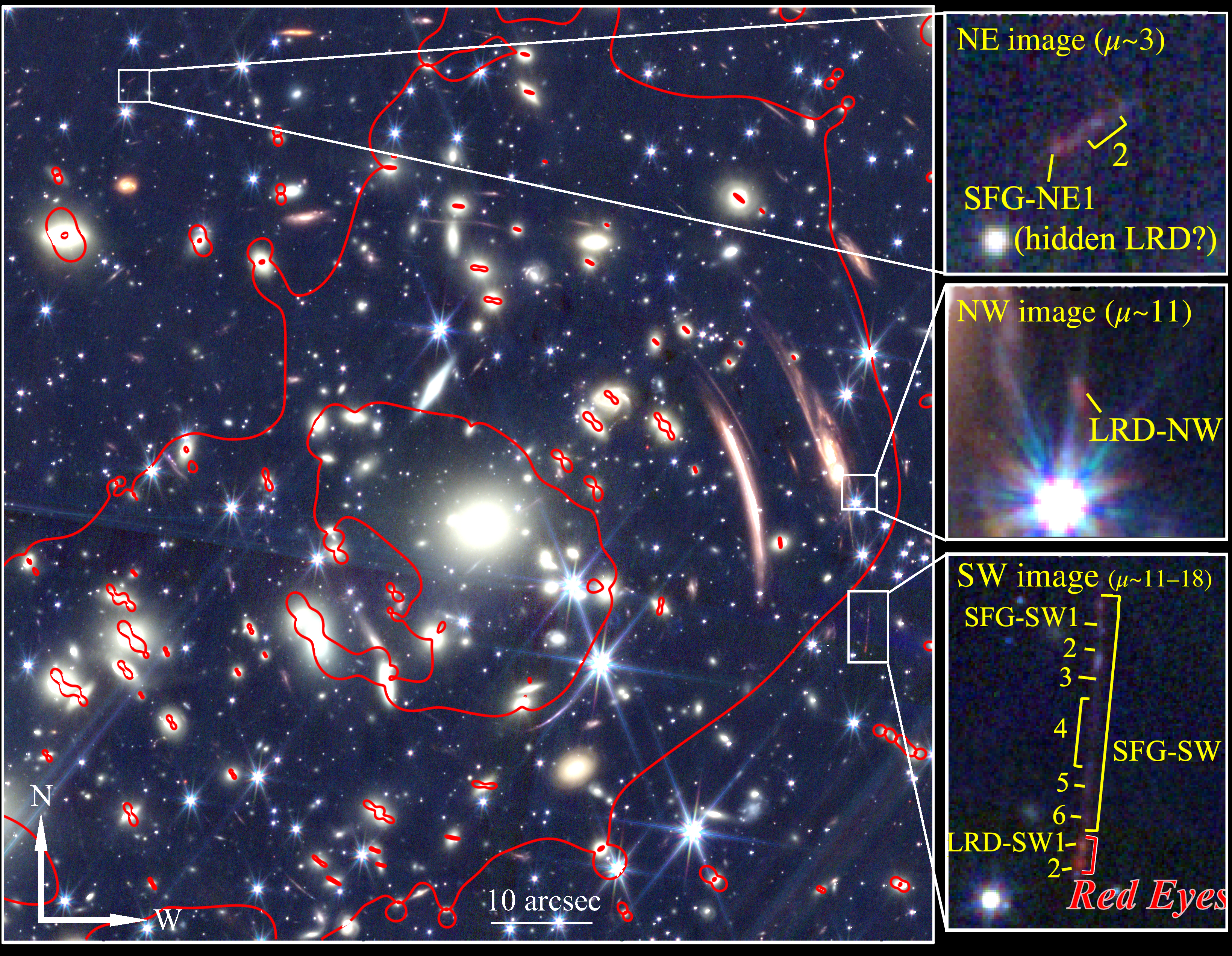}
    \caption{RGB composite image (R: F356W+F444W, G: F200W+F277W, B: F115W+F150W) of the PLCK {G004.5-19.5} cluster. The red contours indicate the critical curves at $z=7.4$. LRD and SFG images are highlighted by annotated inset figures.}
    \label{fig:colorimg}
\end{figure*}

\begin{figure*}
    \centering
    \includegraphics[width=0.8\linewidth]{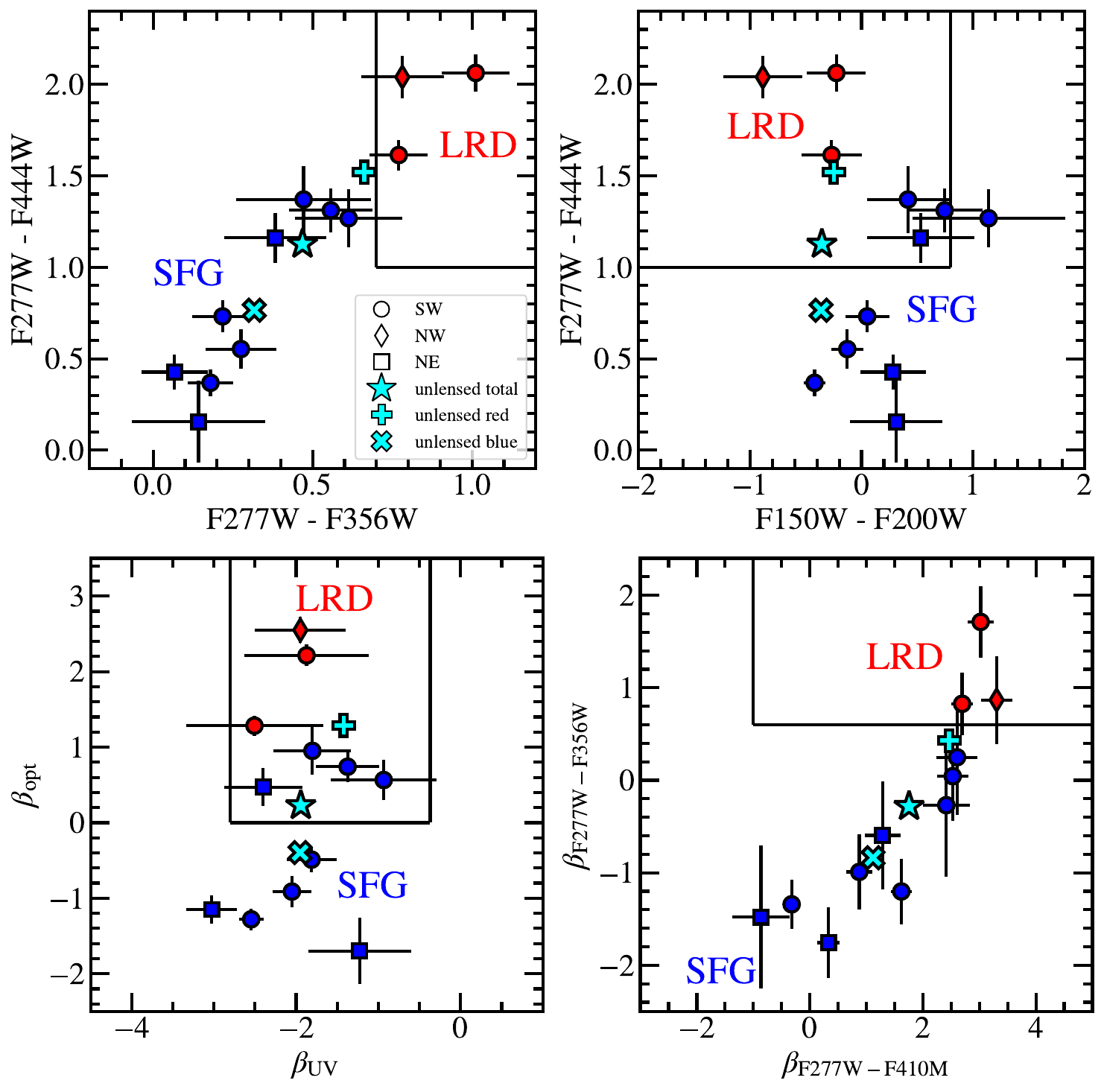}
    \caption{LRD selection using colors (the top panels) and slope fitting (the bottom panels). The black lines show the LRD selection threshold. The red and blue symbols indicate the LRD and SFG images, respectively. The circles, diamonds, and squares represent the SW, NW, and NE images, respectively. { The cyan stars show the unlensed total photometry of the whole system. The cyan plus and X symbols show $0.3''$-diameter aperture photometry centered on the red and blue peaks of the unlensed best-fit model, respectively (``unlensed red'' and ``unlensed blue'').} The LRD-SW1, SW2, and NW satisfy all of the LRD selection criteria, while the SFG components and unlensed mock observed image do not.}
    \label{fig:colorselection}
\end{figure*}

\section{Observations and Data Reduction}\label{sec:obs}
The galaxy cluster PLCK G004.5-19.5 ($z = 0.54$; \citealp{Sifon+2014, Albert+2017, Tarrio+2019, Salmon+2020}) was observed with JWST/NIRCam as part of the treasury lensing cluster survey Vast Exploration for Nascent, Unexplored Sources (VENUS; Cycle 4 GO-6882; PIs: S. Fujimoto \& D. Coe; Fujimoto et al. in prep.) on 2025 October 4. The NIRCam observations utilize ten filters of F090W, F115W, F150W, F200W, F210M, F277W, F300M, F356W, F410M, and F444W with exposure times of $0.35$--$0.57$ hours to homogeneously achieve a source detection limit of $\sim28$ mag across all NIRCam filters (5$\sigma$, point source).
We also make use of archival HST ACS data taken in the RELICS survey ({GO-14096}; PI: D. Coe; \citealp{Coe+2019}). PLCK G004.5-19.5 was observed on 2016 August 15th with F814W and on 2016 September 25th with F606W, with exposure times of $2377\,\mathrm{s}$ and $2180\,\mathrm{s}$, achieving source detection limits of 27.1 and 27.6 mag, respectively.

Here, we briefly explain our data reduction procedure. We begin with JWST level-2 products from MAST and reduce them with the \texttt{grizli} pipeline \citep{Brammer_2021, Brammer_2023}, which is similar to the method used for the widely available public DJA products. The photometric calibration was carried out with the Calibration Reference Data System (CRDS) context file \texttt{jwst\_1456.pmap}. The \texttt{grizli} procedure incorporates crucial improvements over the standard STScI pipeline, including corrections for cosmic rays, stray light and detector artifacts \citep{Bradley+2023, Rigby+2023}. We further incorporate additional background, $1/f$ noise and diffraction spike subtraction, both at the amplifier level, for each filter, and then to the final drizzled mosaic (see, e.g., \citealp{Endsley+2024, Kokorev+2025}). JWST NIRCam images are then drizzled to a 0\farcs03 
/pix grid. The HST images are based on Gaia-aligned mosaics from the CHArGE archive \citep{Kokorev+2022}, which are drizzled on the same footprint and pixel scale as the JWST data. A reduced color composite image is shown in Figure \ref{fig:colorimg}.
More details of the data reduction will be presented in a separate paper.

\begin{figure*}
    \centering
    \includegraphics[width=0.9\linewidth]{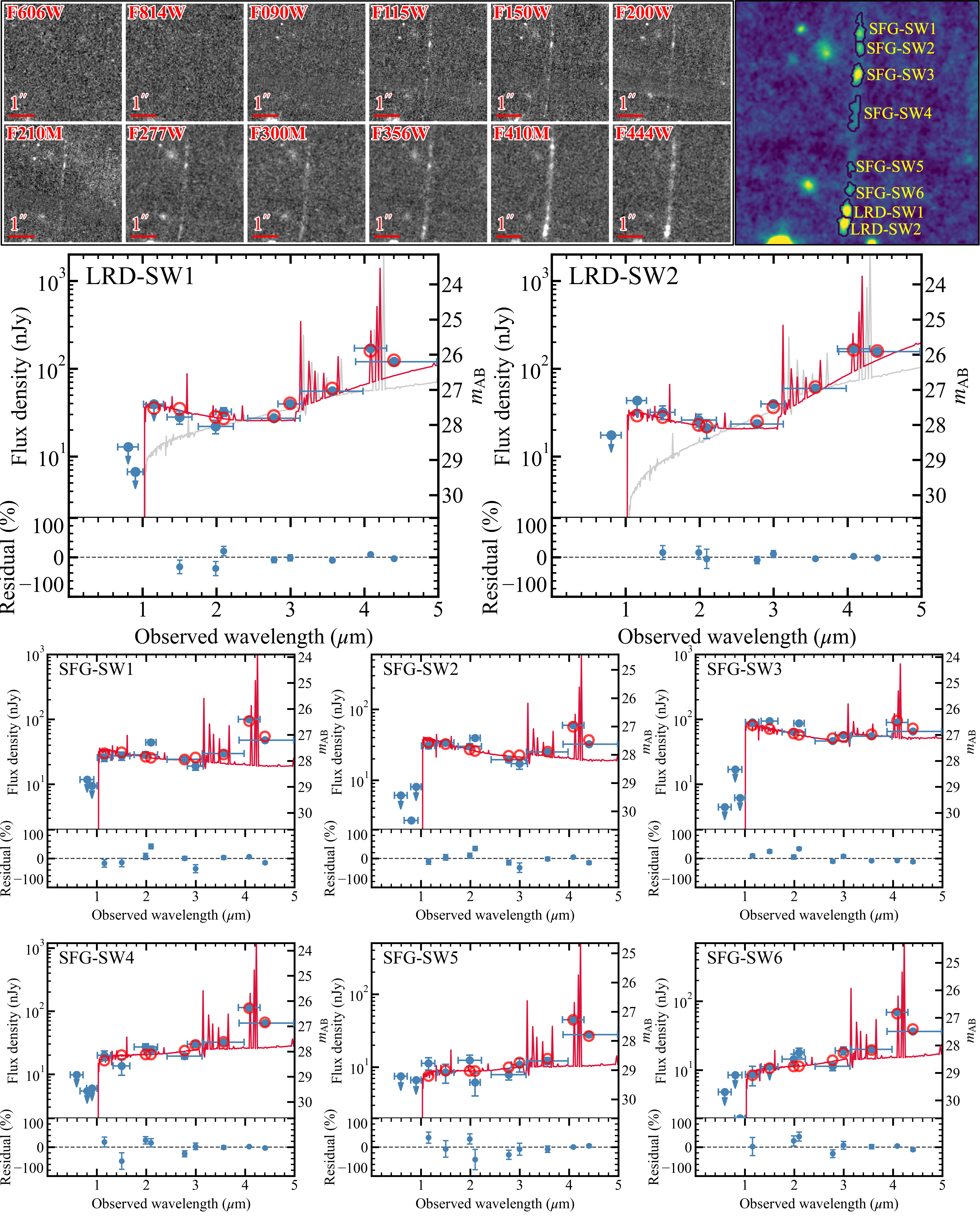}
    \caption{(Top) HST and NIRCam cutouts of the SW image. The rightmost panel represents the segmentation image. (Bottom) Photometry and best-fit SEDs for the SW images. The blue points indicate the observed photometry, with upper limits shown at the $2\sigma$ level. The photometry and upper limit values are not corrected for the magnification. The red curves show the SEDs consisting of galaxy and blackbody components, while the red circles indicate the model photometry. The gray curves in the bottom two panels represent the best-fit SEDs without the blackbody component. The models without the blackbody component favor the solutions invoking dusty SFGs, which significantly underestimate the UV continuum due to the strong dust attenuation. { The panels beneath the SEDs show the percentage residuals, $100(F_{\rm obs}-F_{\rm model})/F_{\rm obs}$. }}
    \label{fig:sed-sw}
\end{figure*}

\begin{figure}
    \centering
    \includegraphics[width=1\linewidth]{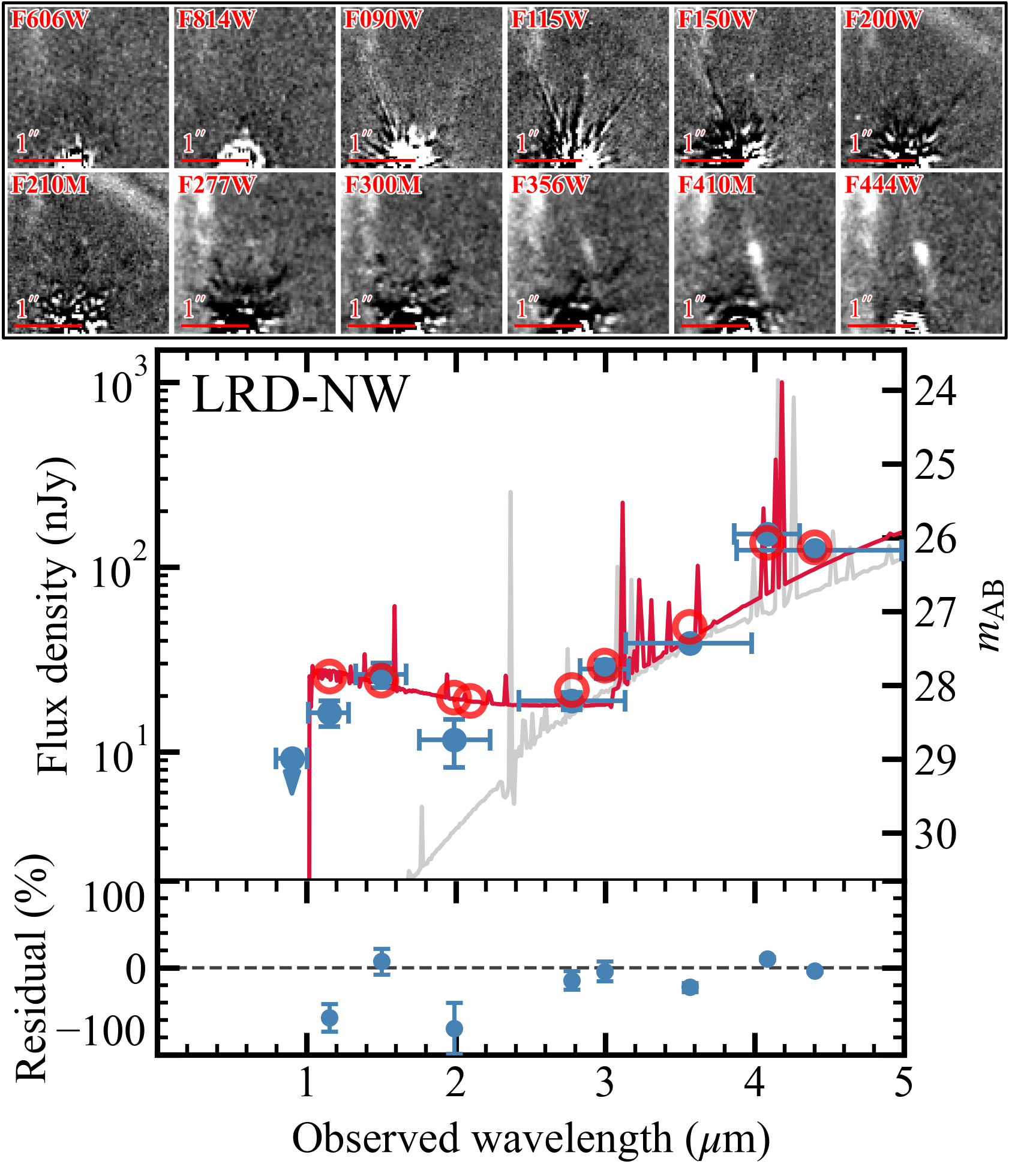}
    \caption{Same as Figure \ref{fig:sed-sw}, but for the NW image.}
    \label{fig:sed-nw}
\end{figure}

\begin{figure}
    \centering
    \includegraphics[width=1\linewidth]{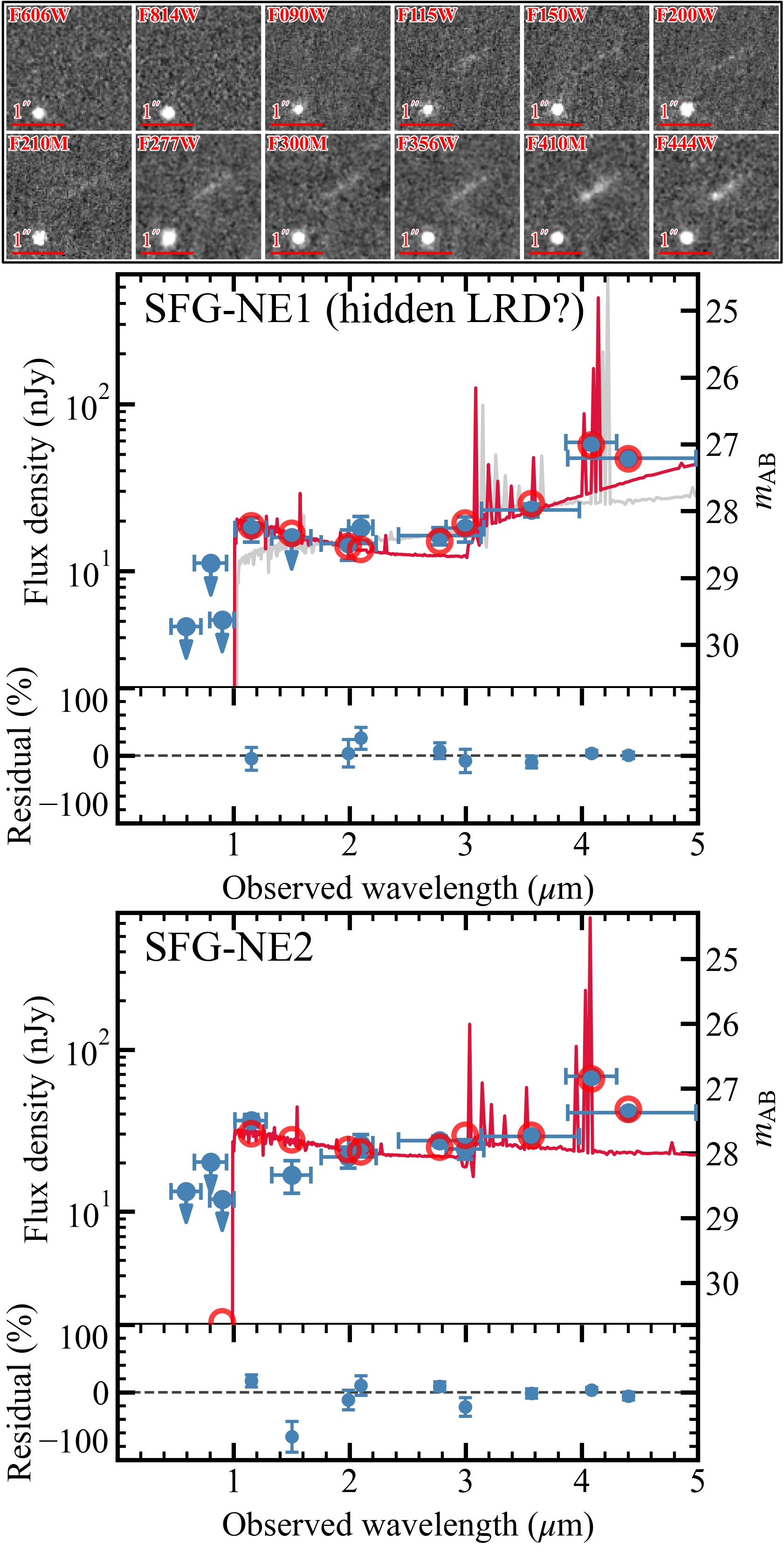}
    \caption{Same as Figure \ref{fig:sed-sw}, but for the NE image. }
    \label{fig:sed-ne}
\end{figure}

\section{Analysis}\label{sec:analysis}
\subsection{Photometry and LRD Selection}

We produce a detection image by taking an inverse variance-weighted mean of the point spread function (PSF)-matched images of F200W, F277W, F356W, and F444W. Source detection is conducted on the detection image using \texttt{Photutils} \citep{Bradley+2025}. 
We then measure isophotal fluxes to obtain reliable SEDs for objects with complex morphology due to the strong gravitational lensing. We measure the total flux of each filter with an isophotal aperture, which is determined by a region $5\sigma$ above the background with the detection image. 

Using the photometry measured above, we select objects that meet the LRD selection criteria presented in \cite{Labbe+2025}:

\begin{align}\label{eq:lrd_selection}
    \mathrm{F150W - F200W} &< 0.8 \\
    \mathrm{F277W - F356W} &> 0.7 \\
    \mathrm{F277W - F444W} &> 1.0,
\end{align}
which selects objects with blue UV and red optical continuum at $z=6-9$ (the top panels of Figure \ref{fig:colorselection}). We do not include the compactness criterion to avoid excluding LRDs with complex morphologies, such as those with extended host galaxies or closely separated multiple LRDs.  
Among the objects satisfying the above criteria, we identify LRD-SW1 and LRD-SW2, which are located near the end of an arc extending $\sim5^{\prime\prime}$ 
(Figure \ref{fig:colorimg}). Hereafter, we refer to LRD-SW1 and 2 as ``Red Eyes." While Red Eyes show clear V-shaped SEDs, which is a typical feature of LRDs, the rest of the arc does not show red optical continua (Figure \ref{fig:sed-sw}) but instead shows an SED typical of star-forming galaxies (SFGs). Red Eyes and the arc all present clear non-detection blueward of F115W and strong excess in F410M, indicating that they are located at the same redshift, and that the arc is an SFG associated with Red Eyes. We thus refer to the arc as SFG-SW. SFG-SW consists of six diffuse components, aligned from north to south and labeled SFG-SW1 through SFG-SW6. 
Based on a gravitational lens model (Section \ref{sec:lensmodel}), we also identify a counter-images in north-west and north-east region (Figure \ref{fig:colorimg}; hereafter NW and NE images, respectively), which have similar colors to Red Eyes. Because the NW image appears to be contaminated by the light from a nearby bright star, we fit the PSF of each band to the star and subtract it before performing photometry. We then identify LRD-NW, which also satisfies the color selection criteria (Figure \ref{fig:colorselection}), although the detection of a galaxy near the NW image remains marginal due to the nearby bright star. There are two distinct extended sources in the NE image, which do not satisfy the color selection criteria (Figure \ref{fig:colorselection}). They are hereafter referred to as SFG-NE1 and 2, although SFG-NE1 shows a mildly V-shaped SED. A possible explanation for the less red optical continuum of SFG-NE1 is presented in Section \ref{sec:sourceplane}. 

We also perform power-law fitting to the UV and optical slopes to see whether these objects satisfy the selection criteria presented in \cite{Kocevski+2025}:

\begin{align}\label{eq:lrd_selection}
    -2.8<\beta_\mathrm{UV} &< -0.37 \\
    \beta_\mathrm{opt} &> 0,
\end{align}
where $\beta_\mathrm{UV}$ and $\beta_\mathrm{opt}$ are the UV and optical slopes, respectively. $\beta_\mathrm{UV}$ is measured with F115W, F150W, and F200W, whereas $\beta_\mathrm{opt}$ is measured with F277W, F356W, and F444W. The bottom left panel of Figure \ref{fig:colorselection} shows $\beta_\mathrm{UV}$ and $\beta_\mathrm{opt}$ values for the LRD and SFG images. We find that all LRD images satisfy the slope selection criteria within the error. 
\cite{Kocevski+2025} introduced additional selection criteria for objects with $z<8$ to exclude emission line galaxies using $\beta_\mathrm{F277W-F356W}$, and $\beta_\mathrm{F277W-F410M}$, which are the power-law slopes measured with the two bands described in the subscripts. We use 

\begin{align}\label{eq:lrd_selection_revised}
    \beta_\mathrm{F277W-F356W} &> 0.6 \\
    \beta_\mathrm{F277W-F410M} &> -1,
\end{align}
where the lower limit on $\beta_\mathrm{F277W-F356W}$ is equivalent to the color cut $\mathrm{F277W}-\mathrm{F356W} > 0.7$. 
The bottom right panel of Figure \ref{fig:colorselection} shows $\beta_\mathrm{F277W-F356W}$ and $\beta_\mathrm{F277W-F410M}$. We find that all LRD images satisfy these criteria, indicating that the red optical slopes are not solely the result of strong emission lines. Several SFG components satisfy a subset of the selection criteria, but not all of them. Although these components exhibit slightly red optical colors, their SEDs can be explained by emission-line galaxies (Section \ref{sec:results_sed}).

\subsection{SED Fitting}\label{sec:sed_fitting}
We conduct SED fitting to the LRD and SFG images using \texttt{BAGPIPES} \citep{Carnall+2018}. 
For stellar emission, we assume a BPASS stellar population synthesis model \citep{Stanway_Eldridge_2018} 
with delayed-$\tau$ star-formation histories (SFHs). 
We adopt flat priors on stellar age $t_\mathrm{age} \in [0.01,5]\,\mathrm{Gyr}$, star-formation timescale $\tau \in [0.1,3]\,\mathrm{Gyr}$, stellar metallicity $Z_* \in [0.1,1]\,\mathrm{Z_\odot}$, and {log of the total formed stellar mass} ${\log_{10}(M_\mathrm{formed}/M_\odot) \in [5,13]}$.
For nebular emission, we run version 23 of the \texttt{Cloudy} photoionization code \citep{Ferland+1998, Gunasekera+2023} using the BPASS spectra as incident radiation and obtain grids of nebular emission. We adopt a flat prior on the ionization parameter $\log_{10}U \in [-4,-1]$, and the nebular metallicity is assumed to be the same as the stellar metallicity. For dust extinction, we assume a \cite{Calzetti+1994} extinction law with a flat prior on the total extinction $A_\mathrm{V} \in [0, 5]\,\mathrm{mag}$. The redshift prior is set to ${z\in[0,9]}$. In addition to the galaxy model fitting explained above, to model the red optical continuum observed in the LRDs, we follow the method described in \cite{deGraaff+2025b} and \cite{Umeda+2025} and include a blackbody function:

\begin{equation}
    f_\nu \propto B_\nu(T_\mathrm{BB})H(x),
\end{equation}
where $B_\nu(T_\mathrm{BB})$ is the Planck function of temperature $T_\mathrm{BB}$. We take into account the Balmer break by introducing the Heaviside step function $H(x)$, where ${x =c/\lambda_\mathrm{Balmer} - \nu}$ and $\lambda_\mathrm{Balmer}=3646\,\mathrm{\AA}$. For the LRD fits, we adopt a flat prior of $T_\mathrm{BB}\in[{2000},6000]\,\mathrm{K}$. In the blackbody fitting, we fix $A_\mathrm{V}=0$, motivated by the negligible dust content in the LRDs and by the need to avoid degeneracy with $T_\mathrm{BB}$. We also impose a constraint on the blackbody fit such that the Balmer break wavelength is shorter than F356W, in order to prevent an unphysically strong Balmer break from mimicking the red optical continuum. We perform SED fitting with and without blackbody emission to all components. {The best-fit parameters and their uncertainties are listed in Table \ref{tab:sed-fitting}.}

\subsection{Gravitational Lens Modeling}\label{sec:lensmodel}
A gravitational lens model for the PLCK {G004.5-19.5} cluster is constructed based on a revised version of the parametric method by \citealp{Zitrin+2015} (for more details, see also \citealp{Pascale+2022}; \citealp{Furtak+2023_lensing}). In this model, cluster member galaxies are described by double Pseudo-Isothermal Ellipsoids (dPIE; \citealp{Eliasdottir+2007}), and the large-scale dark matter distribution is represented by a diffuse halo modeled as a Pseudo-Isothermal Elliptical Mass Distribution (PIEMD; e.g. \citealp{Keeton+2001}). We use {9} multiple image systems as constraints, {two of which are distinct clumps within the same lensed galaxy}. {Four systems are spectroscopically confirmed} for at least one image. Cluster members are selected by matching the red sequences formed in the NIRCam/F090W and F150W; ACS/F814W and NIRCam/F150W; and ACS/F606W and F814W spaces, together with data from MUSE. For minimization, we adopt a positional uncertainty of 0\farcs1 for all images. All multiple images are reproduced by the model with a $\Delta$RMS between model and observations of 0\farcs57. 
{The best-fit magnification factors and their uncertainties are listed in Table \ref{tab:sed-fitting}.}


\begin{table*}[t]
\caption{{BAGPIPES SED-fitting results for the image-plane components.}}
\label{tab:sed-fitting}
\centering
\begingroup

\scriptsize
\renewcommand{\arraystretch}{1.15}
\setlength{\tabcolsep}{2.5pt}
\hspace*{-67pt}%
\resizebox{\textwidth}{!}{%
\begin{tabular}{lcccccc}
\hline\hline
Component & R.A. (deg) & Decl. (deg) & $\mu$ & $z$ & $\log(M_{\star,\mathrm{int}}/M_\odot)$ & \shortstack{$L_{\mathrm{BB,int}}$\\($10^{42}\,\mathrm{erg\,s}^{-1}$)} \\
\hline
LRD-SW1 & 289.25876280 & $-33.52560804$ & ${11.4^{+4.5}_{-0.9}}$ & ${7.41^{+0.03}_{-0.04}}$ & ${6.36^{+0.04}_{-0.15}}$ & ${5.0^{+1.0}_{-1.5}}$ \\
LRD-SW2 & 289.25877523 & $-33.52567500$ & ${11.1^{+4.4}_{-1.4}}$ & ${7.38^{+0.04}_{-0.05}}$ & ${6.28^{+0.07}_{-0.15}}$ & ${15.0^{+3.4}_{-4.5}}$ \\
LRD-NW  & 289.25903160 & $-33.52141160$ & ${11.8^{+3.2}_{-2.7}}$ & ${7.35^{+0.04}_{-0.05}}$ & ${6.18^{+0.12}_{-0.11}}$ & ${11.9^{+4.2}_{-2.9}}$ \\
SFG-NE1 & 289.28256820 & $-33.51046131$ & $3.37^{+0.57}_{-0.18}$ & ${7.27^{+0.09}_{-0.10}}$ & ${7.00^{+0.26}_{-0.19}}$ & ${6.1^{+1.5}_{-1.6}}$ \\
SFG-NE2 & 289.28248058 & $-33.51040913$ & $3.38^{+0.67}_{-0.15}$ & ${7.13^{+0.11}_{-0.08}}$ & ${7.58^{+0.13}_{-0.13}}$ & -- \\
SFG-SW1 & 289.25867600 & $-33.52465074$ & ${18.7^{+6.8}_{-5.2}}$ & ${7.48^{+0.01}_{-0.02}}$ & ${6.38^{+0.17}_{-0.17}}$ & -- \\
SFG-SW2 & 289.25867676 & $-33.52473083$ & ${17.6^{+9.8}_{-3.5}}$ & ${7.48^{+0.02}_{-0.03}}$ & ${6.58^{+0.13}_{-0.21}}$ & -- \\
SFG-SW3 & 289.25869263 & $-33.52487806$ & ${16.2^{+7.4}_{-3.8}}$ & ${7.28^{+0.29}_{-0.06}}$ & ${7.26^{+0.12}_{-0.17}}$ & -- \\
SFG-SW4 & 289.25871090 & $-33.52511145$ & ${14.3^{+4.6}_{-2.9}}$ & ${7.44^{+0.03}_{-0.04}}$ & ${6.95^{+0.15}_{-0.17}}$ & -- \\
SFG-SW5 & 289.25873049 & $-33.52536466$ & ${12.7^{+4.8}_{-1.7}}$ & ${7.44^{+0.06}_{-0.13}}$ & ${6.53^{+0.22}_{-0.22}}$ & -- \\
SFG-SW6 & 289.25874650 & $-33.52550730$ & ${11.9^{+5.2}_{-1.1}}$ & ${7.44^{+0.04}_{-0.06}}$ & ${6.84^{+0.18}_{-0.24}}$ & -- \\
\hline
\end{tabular}%
}

\vspace{0.7em}
\hspace*{-67pt}%
\resizebox{\textwidth}{!}{%
\begin{tabular}{lcccccc}
\hline\hline
Component & $T_\mathrm{BB}$ (K) & $t_\mathrm{age}$ (Myr) & $\tau$ (Gyr) & $Z_\star/Z_\odot$ & $\log U$ & $A_V$ (mag) \\
\hline
LRD-SW1 & ${4580^{+280}_{-280}}$ & ${10.3^{+0.6}_{-0.2}}$ & ${1.6^{+1.0}_{-1.0}}$ & ${0.21^{+0.02}_{-0.01}}$ & ${-1.84^{+0.18}_{-0.13}}$ & 0 (fixed) \\
LRD-SW2 & ${3730^{+150}_{-150}}$ & ${11.0^{+1.7}_{-0.7}}$ & ${1.6^{+0.9}_{-1.0}}$ & ${0.24^{+0.07}_{-0.04}}$ & ${-1.95^{+0.17}_{-0.13}}$ & 0 (fixed) \\
LRD-NW  & ${3610^{+150}_{-140}}$ & ${10.2^{+0.4}_{-0.2}}$ & ${1.6^{+0.9}_{-1.0}}$ & ${0.21^{+0.02}_{-0.01}}$ & ${-1.75^{+0.17}_{-0.17}}$ & 0 (fixed) \\
SFG-NE1 & ${4410^{+490}_{-290}}$ & ${54^{+68}_{-24}}$ & ${1.6^{+0.9}_{-0.9}}$ & ${0.32^{+0.16}_{-0.11}}$ & ${-1.85^{+0.31}_{-0.26}}$ & 0 (fixed) \\
SFG-NE2 & -- & ${120^{+61}_{-36}}$ & ${1.7^{+0.9}_{-0.9}}$ & ${0.34^{+0.16}_{-0.11}}$ & ${-1.55^{+0.30}_{-0.27}}$ & ${0.11^{+0.08}_{-0.07}}$ \\
SFG-SW1 & -- & ${14.4^{+5.6}_{-3.0}}$ & ${1.6^{+0.9}_{-1.0}}$ & ${0.36^{+0.14}_{-0.12}}$ & ${-1.37^{+0.26}_{-0.30}}$ & ${0.19^{+0.09}_{-0.08}}$ \\
SFG-SW2 & -- & ${41^{+16}_{-10}}$ & ${1.7^{+0.8}_{-1.0}}$ & ${0.71^{+0.11}_{-0.11}}$ & ${-1.18^{+0.13}_{-0.22}}$ & ${0.04^{+0.05}_{-0.03}}$ \\
SFG-SW3 & -- & ${123^{+14}_{-21}}$ & ${1.5^{+0.9}_{-0.7}}$ & ${0.98^{+0.02}_{-0.03}}$ & ${-1.07^{+0.05}_{-0.09}}$ & ${0.02^{+0.02}_{-0.01}}$ \\
SFG-SW4 & -- & ${25.3^{+9.7}_{-5.7}}$ & ${1.6^{+1.0}_{-1.0}}$ & ${0.32^{+0.12}_{-0.08}}$ & ${-1.48^{+0.25}_{-0.30}}$ & ${0.74^{+0.11}_{-0.10}}$ \\
SFG-SW5 & -- & ${21^{+16}_{-8}}$ & ${1.6^{+0.9}_{-1.0}}$ & ${0.32^{+0.17}_{-0.09}}$ & ${-1.36^{+0.24}_{-0.31}}$ & ${0.66^{+0.19}_{-0.15}}$ \\
SFG-SW6 & -- & ${24^{+15}_{-10}}$ & ${1.7^{+0.9}_{-1.0}}$ & ${0.39^{+0.20}_{-0.14}}$ & ${-1.68^{+0.40}_{-0.32}}$ & ${0.85^{+0.16}_{-0.15}}$ \\
\hline
\end{tabular}%
}

\vspace{0.5em}
\parbox{0.98\textwidth}{\footnotesize {Note. The SED-fitting values are posterior medians with uncertainties from the 16th and 84th percentiles. The listed magnification factors are from the fiducial lens model, with uncertainties spanning the 16th--84th percentiles of the lens-model realizations. The intrinsic stellar masses and blackbody luminosities are corrected using the component-wise fiducial magnifications listed here. Their quoted uncertainties include the independent SED-fitting and magnification uncertainties. The LRD and SFG-NE1 fits include a blackbody component and use $A_V=0$ fixed. The SFG-NE2 and SFG-SW galaxy-only fits use free $A_V$ and have no $T_\mathrm{BB}$ or $L_\mathrm{BB}$.}}
\endgroup
\end{table*}

\begin{figure*}
    \centering
    \includegraphics[width=1\linewidth]{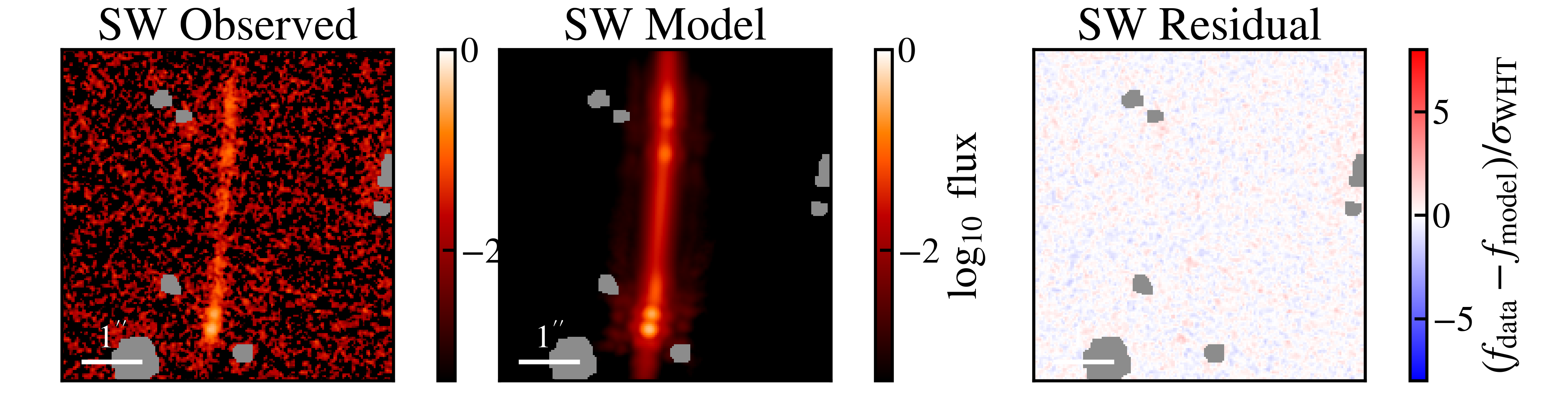}
    \caption{Source plane reconstruction. 
    From left to right, each column represents the observed, modeled, residual of the F444W image. {Gray regions are masked in the fitting. }}
    \label{fig:reconstruction}
\end{figure*}

\begin{figure*}
    \centering
    \includegraphics[width=0.7\linewidth]{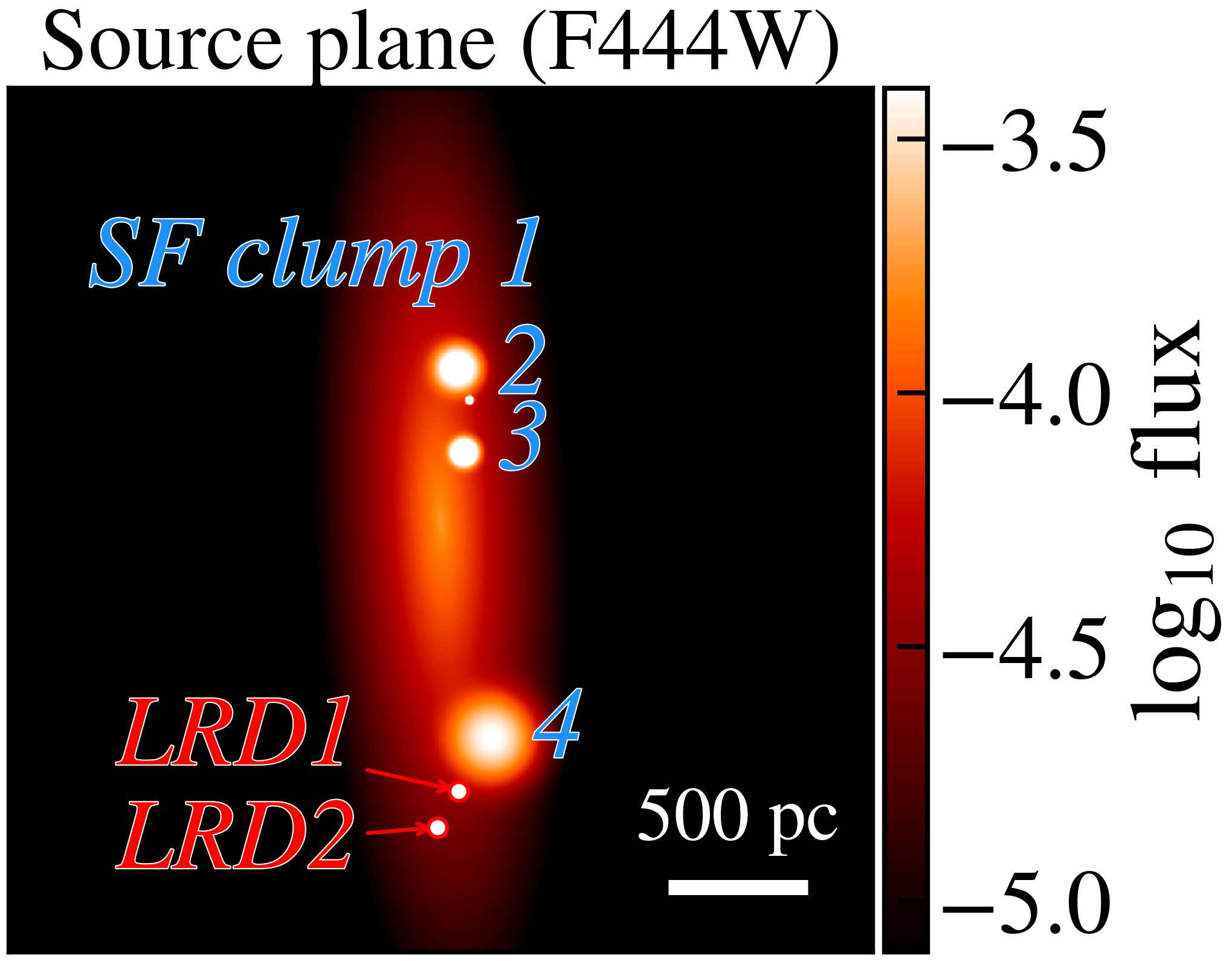}
    \caption{Source plane image. The scale bar indicates 500 physical pc at $z=7.4$.}
    \label{fig:sourceplane}
\end{figure*}

\begin{figure}
    \centering
    \includegraphics[width=1\linewidth]{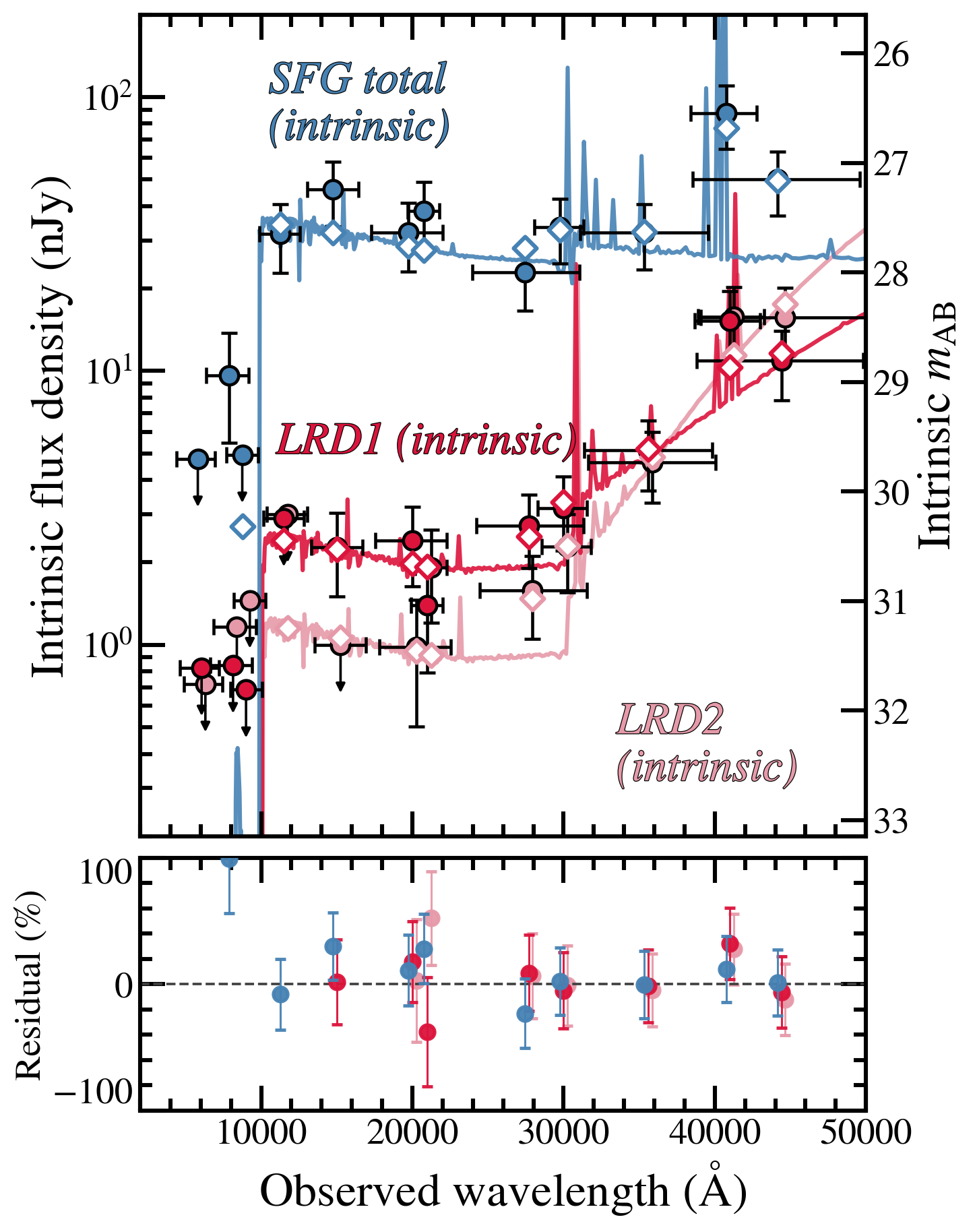}
    \caption{Intrinsic SEDs obtained by multi-band source plane reconstruction (Section \ref{sec:sourceplane}). The blue points indicate the SED for the SFG (sum of all of the star-forming clumps and diffuse component), while the red and pink points represent that for each of the LRD pair. { The solid curves show the maximum-likelihood \texttt{BAGPIPES} models, and the open diamonds show the corresponding model photometry in each filter. Non-detections are displayed as $2\sigma$ upper limits in the upper panel. The lower panel shows the percentage residuals, $100(F_\mathrm{obs}-F_\mathrm{model})/F_\mathrm{obs}$.}}
    \label{fig:sed-int}
\end{figure}

\section{Results and Discussion}\label{sec:results}
\subsection{SEDs of Individual Component}\label{sec:results_sed}
Figures \ref{fig:sed-sw}, \ref{fig:sed-nw}, and \ref{fig:sed-ne}, show the SED-fitting results for each component. The photometric redshifts of the individual components are generally consistent at $z \sim 7.4$, {ranging from $z=7.1$ to $z=7.6$. This range is consistent with the redshift range where the H$\beta$+[O \textsc{iii}]$\lambda\lambda$4959,5007 emission lines fall in F410M ($\sim3.9$--4.3$\mu \mathrm{m}$).} The \texttt{BAGPIPES} fits that include a blackbody component reproduce both the blue UV continuum and the red optical continuum observed in the LRD images. Without this component, the best-fit solutions require dusty star-forming galaxy templates to match the red optical continuum, which in turn leads to a significant underestimate of the UV flux. In contrast, the SFG components are well-explained by young star-forming galaxy models. We note that the fits to the SFG images 
with and without the blackbody component yield nearly identical best-fit SEDs, implying little contribution from the blackbody component. 
{ For the LRD images, the inferred $T_\mathrm{BB}$ values span $\sim3600$--$4600\,\mathrm{K}$, which are lower than theoretical predictions of $\sim5000\,\mathrm{K}$ from the black hole envelope models \citep{Kido+2025, Inayoshi+2025} but consistent with several LRDs identified in e.g., \citet{deGraaff+2025b,Wang+2026}.}


\subsection{Source Plane Reconstruction}\label{sec:sourceplane}
Using the gravitational lens model established in Section \ref{sec:lensmodel}, we reconstruct a source plane image. We utilize \texttt{Lenstruction} \footnote[1]{\url{https://github.com/ylilan/lenstruction}} \citep{Yang+2020}, which is built based on \texttt{Lenstronomy} \footnote[2]{\url{https://github.com/lenstronomy/lenstronomy}} \citep{Birrer+2018, Birrer+2021}. \texttt{Lenstruction} performs forward modeling to reconstruct the source brightness distribution, considering the broadening due to the PSF. 

As shown in Figure \ref{fig:colorimg}, the critical curve does not cross between the LRD-SW1 and 2. This indicates that the two LRD images are not multiple images of a single LRD, but instead represent two distinct LRDs on the source plane. 
While they are not fully resolved in the NE and NW images owing to their relatively low magnifications {and nearby star contamination, respectively,} they are clearly deblended in the SW image\footnote[3]{If we consider a small perturber in between LRD-SW1 and 2, this could split a single LRD image into multiple images. At this position relative to the critical curve, such a perturber would produce three counter-images very close to each other, not two. 
For a sufficiently small perturber, two of the images could merge into a single one, leaving only two visible images as seen in the JWST data. However, in such a scenario, the flux of one of the counter-images (the unresolved pair) would be significantly higher than the other, which is inconsistent with the observations. Furthermore, there is no evidence of the perturber in the images, making this explanation highly unlikely.}.

{ We thus use two point-source components to describe the two LRD images. To reproduce the SFG components, we model the diffuse emission covering the entire arc with a single elliptical \sersic profile and the clumpy emission (SFG-SW1, 2, 3, and 6) with four Gaussian profiles, allowing the widths of the Gaussian profiles to vary freely.} To avoid a degeneracy, the \sersic index $n$ is fixed at 1. The fitting is conducted to the F444W image of the SW image, which have the high S/N ratio (multi-band fitting is conducted later in this section).
The result is presented in Figure \ref{fig:reconstruction}. { The two point-source components successfully explain LRD-SW1 and 2.} The intrinsic two LRDs corresponding to LRD-SW1 and LRD-SW2 are hereafter referred to as LRD1 and LRD2, respectively. { The four Gaussian SFG components and the \sersic profile reproduce the clumpy morphology embedded within the diffuse arc.} 
The source plane image is shown in Figure \ref{fig:sourceplane}. LRD1 and LRD2 are separated by a distance of $\sim {150}\,\mathrm{pc}$ (physical), which is much smaller than those of known dual LRDs (a few kpc; \citealp{Tanaka+2024, Merida+2025}). The two LRDs are located at a distance of $\sim {900}\,\mathrm{pc}$ from the center of the SFG, which is similar to several offsets reported in previous LRD studies, such as blue companions observed near LRDs (\citealp[e.g.,][]{Golubchik+2025, Baggen+2025}, {\citealp{Barger+2026}}) and off-centered diffuse emission around LRDs \citep{CChen+2025a, Rinaldi+2025}. The half light radius of the SFG is $\sim {475}\,\mathrm{pc}$, which corresponds to the typical size of star-forming galaxies at this redshift and luminosity \citep[e.g.,][]{Shibuya+2019, Ono+2025, Morishita+2024, Miller+2025}.

Next, we conduct multi-band fitting to investigate the intrinsic SEDs of the LRDs and SFG. 
{ For each band, we fit the widths of the four Gaussian SFG components and all seven non-negative component amplitudes, while fixing the all component positions to the F444W best-fit model. We also fix the \sersic structural parameters (effective radius, axis ratio, and position angle) to the values obtained from the F444W fit, where the diffuse component is best constrained. Allowing both the Sérsic morphology and normalization to vary independently in the lower-S/N bands leads to poorly constrained and unstable fits.} We then integrate the surface brightness distributions of the \sersic and Gaussian profiles to obtain the total flux of each component. { Figure \ref{fig:sed-int} presents the intrinsic SEDs of the two LRDs and the total SFG.} 
Interestingly, the intrinsic luminosity of the LRDs ($M_\mathrm{UV,int}\gtrsim {-17}$) are $\sim20$ times fainter than that of the SFG ($M_\mathrm{UV,int}\sim-19$).

\begin{figure*}
    \centering
    \includegraphics[width=1\linewidth]{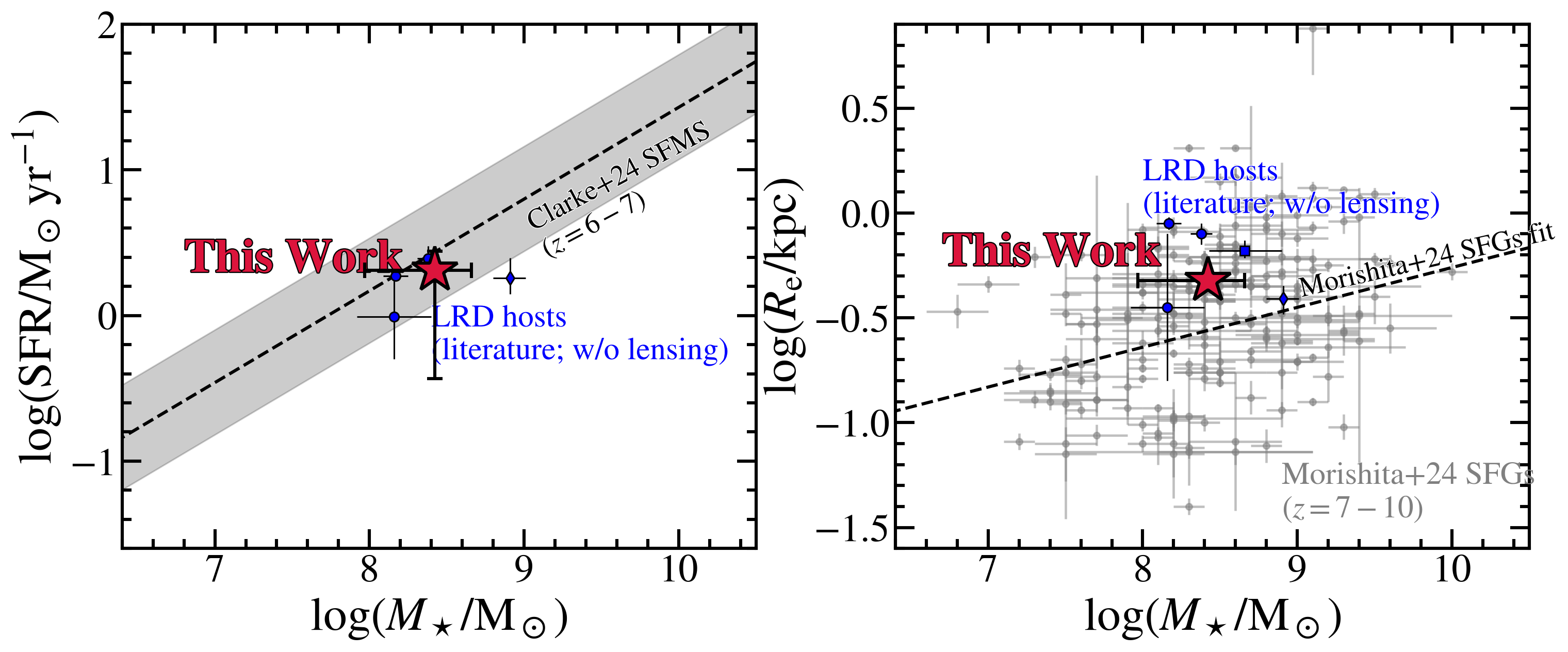}
    \caption{(Left) SFR-stellar mass relation. The red and blue symbols are the same as the left figure. The black dashed line and gray shaded region are the best-fit relation and its uncertainty for SFMS presented in \cite{Clarke+2024}. (Right) Size-mass relation. The red star shows the SFG in this work. The blue points represent the LRD hosts presented in literature (circle: \citealp{CChen+2025a}; square: \citealp{CChen+2025b}; diamond: \citealp{Zhang+2025}.) The gray points and black dashed line indicate SFG sample and linear regression to it presented in \citep{Morishita+2024}.}
    \label{fig:sfg-properties}
\end{figure*}

We conduct the \texttt{BAGPIPES} fitting to the intrinsic SEDs in the same manner as described in Section \ref{sec:sed_fitting}. Figure \ref{fig:sed-int} presents the {maximum-likelihood} SEDs, showing a solution of a young SFG. { The SFG has an intrinsic stellar mass of $\log (M_\star/\mathrm{M_\odot}) = 8.42^{+0.24}_{-0.46}$, a mass-weighted stellar age of ${100^{+120}_{-60}} \,\mathrm{Myr}$, and an SFR of ${2.1^{+0.7}_{-1.7}}\,\mathrm{M_\odot\,yr^{-1}}$.} { We also fit the intrinsic SEDs of LRD1 and LRD2 using the same galaxy-plus-blackbody model adopted for their image-plane SEDs, obtaining $L_{\mathrm{BB,int}}={1.0^{+2.0}_{-0.6}}\times10^{43}$ and ${2.3^{+6.1}_{-1.5}}\times10^{43}\,\mathrm{erg\,s^{-1}}$, respectively. As a sanity check, correcting their image-plane luminosities, $L_{\mathrm{BB,obs}}={5.74^{+0.96}_{-0.76}}\times10^{43}$ and $1.67^{+0.29}_{-0.22}\times10^{44}\,\mathrm{erg\,s^{-1}}$, by the corresponding magnifications, $\mu={11.4^{+4.5}_{-0.9}}$ and ${11.1^{+4.4}_{-1.4}}$, gives $L_{\mathrm{BB,obs}}/\mu={5.0^{+1.0}_{-1.5}}\times10^{42}$ and $1.50^{+0.34}_{-0.45}\times10^{43}\,\mathrm{erg\,s^{-1}}$, respectively, after propagating the independent luminosity and magnification uncertainties in logarithmic space. These values are consistent with those inferred directly from the intrinsic SEDs within the uncertainties.} 
The left panel of Figure \ref{fig:sfg-properties} shows the relation between SFR and stellar mass. The SFG in this work is located on the star-forming main sequence (SFMS), indicating that it corresponds to the typical SFG at this redshift. The right panel of Figure \ref{fig:sfg-properties} presents the size-mass relation, where we find that the host galaxy has a typical size at this redshift and stellar mass. 

{ Using the gravitational lens model, we also constrain the source plane sizes of the LRDs. Two-demensional Gaussian fit to the F444W image, accounting for the empirical PSF and local lensing shear, gives source-plane Gaussian half-light radii of $R_{\rm e}\simeq27$ and $34\,\mathrm{pc}$ for LRD-SW1 and 2, respectively. The finite sizes may partly include host-galaxy light blended with an unresolved nucleus, so these values should be regarded as upper limits for the central engine of LRDs. Nevertheless, both components remain extremely compact and are consistent with an AGN interpretation \citep[e.g.,][]{Furtak+2023, Killi+2024}. }

\subsection{Mock Observation without Gravitational Lensing}
\begin{figure*}
    \centering
    \includegraphics[width=1\linewidth]{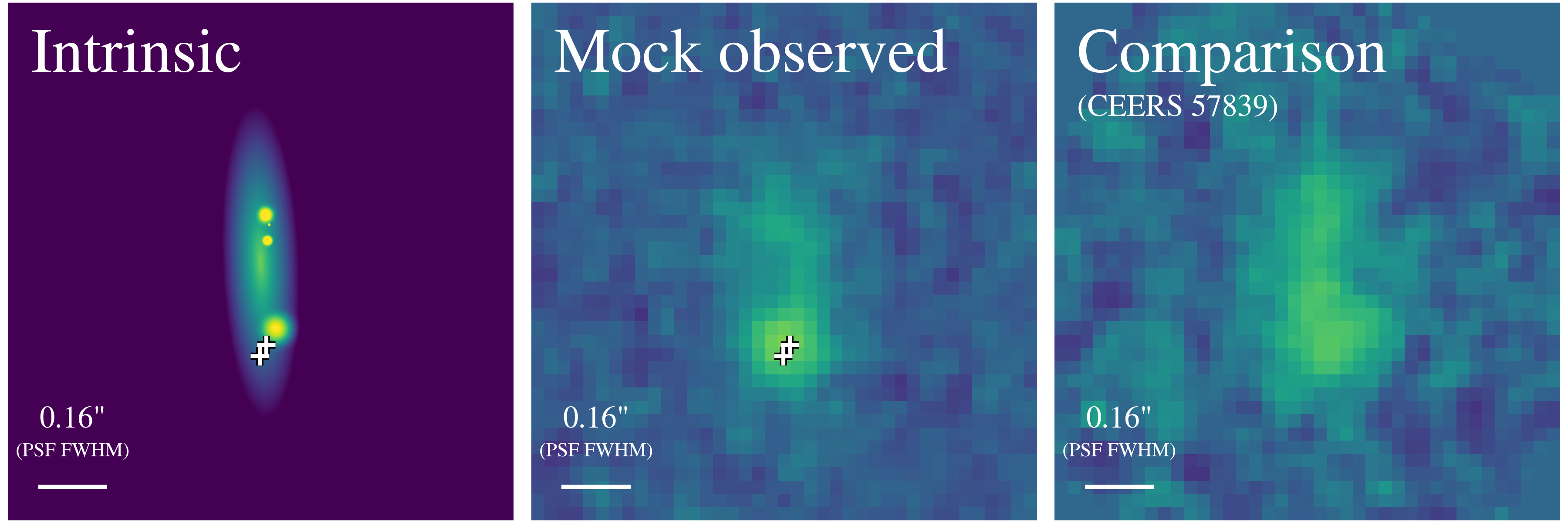}
    \caption{NIRCam F444W mock image without gravitational lensing. (Left) Intrinsic image obtained from the source plane reconstruction (Section \ref{sec:sourceplane}). (Middle) Mock observed image obtained by convolving the NIRCam PSF to the intrinsic image. {The white plus signs in the left and middle panels show the positions of the LRDs.} (Right) Example image of the typical SFG ({CEERS-57839}), which has the similar redshift and luminosity as the SFG in this work.}
    \label{fig:mockobs}
\end{figure*}

Owing to strong gravitational lensing, we are able to observe a pair of faint LRDs associated with an intrinsically dominant, otherwise typical SFG. To investigate how such a system would appear without gravitational lensing, we generate a mock image by convolving the source-plane image with the F444W PSF (Figure \ref{fig:mockobs}). For comparison, we present an example of a typical star-forming galaxy, {CEERS-57839}, which has a similar redshift (${z_\mathrm{phot}=6.6}$; {\citealp{Cox+2026}}) and UV luminosity (${M_\mathrm{UV}=-19.9}$; {\citealp{Ono+2024}}). The mock image resembles typical SFGs at similar redshifts, {and two LRDs are completely blended.} This suggests that we may be missing such faint and multiple LRD systems associated with ordinary SFGs in the absence of strong gravitational lensing. 

\begin{figure*}
    \centering
    \includegraphics[width=1\linewidth]{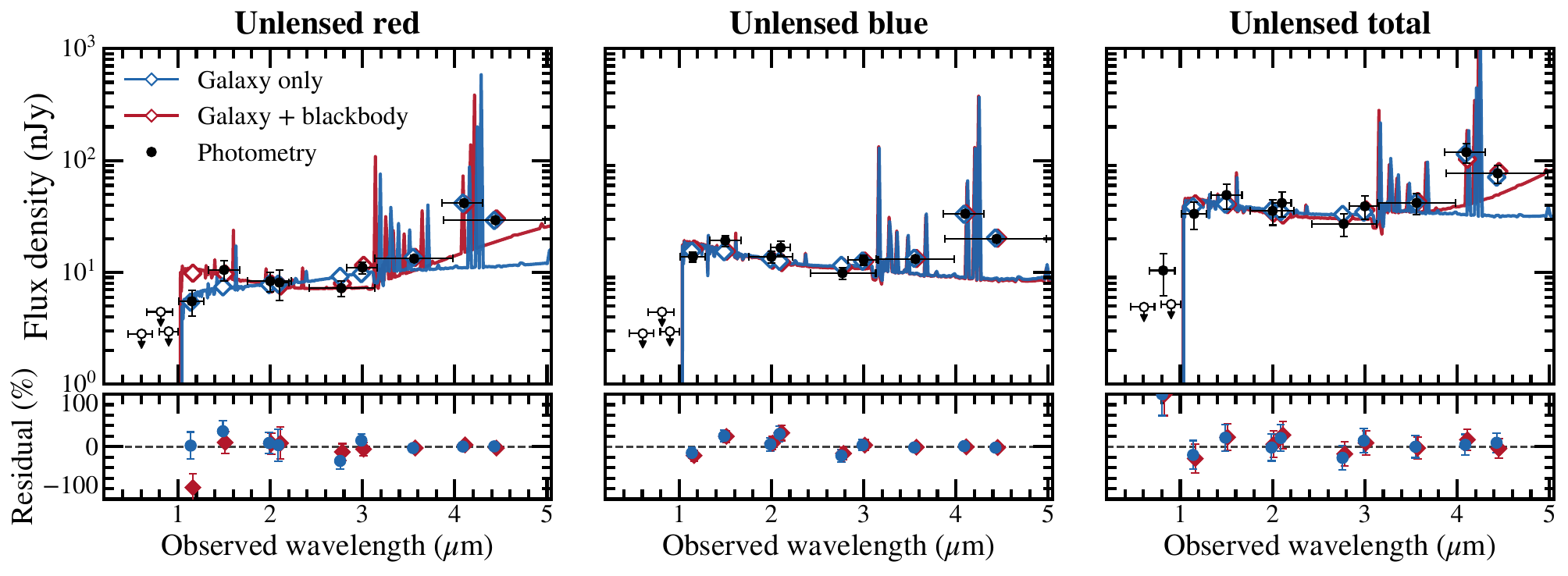}
    \caption{{ SED fitting for the unlensed mock SEDs. (Left) SED measured in a $0.3''$-diameter aperture centered on the LRD pair. (Middle) SED measured in the same aperture centered on the blue-side peak of the unlensed mock image. (Right)  SED of the whole system, obtained by summing the SFG, LRD1, and LRD2 components. The blue and red curves show the maximum-likelihood galaxy-only and galaxy-plus-blackbody models, respectively; open diamonds indicate the model photometry, open circles with arrows show $2\sigma$ upper limits, and the lower panels show the percentage residuals, $100(F_\mathrm{obs}-F_\mathrm{model})/F_\mathrm{obs}$.}}
    \label{fig:sed-unlensed}
\end{figure*}

{ Figure \ref{fig:sed-unlensed} compares SEDs measured in $0.3''$-diameter apertures centered on the LRD pair, the blue-side model peak, and the total SED of the whole system (SFG+LRD1+LRD2). We fit these SEDs with galaxy-only and galaxy-plus-blackbody models described in Section \ref{sec:sed_fitting}. The broadband F444W excess can be reproduced by either a red continuum or strong nebular emission, and does not by itself distinguish between these interpretations. None of the three SEDs, therefore, securely requires a blackbody component. Consistently, the unlensed total, unlensed red, and unlensed blue points all lie outside the LRD selection regions in Figure \ref{fig:colorselection}. This is qualitatively consistent with the fact that the low-magnification NE image does not satisfy the LRD selection criteria.}

{ These results suggest that galaxies hosting similarly faint LRDs may be missed in the absence of gravitational lensing because their integrated colors and morphologies are dominated by the host galaxy. {A potentially related example is the transitional LRD (tLRD) reported by \citet{Merida+2026}, which exhibits compact morphology and broad H$\alpha$ emission but fails the optical LRD selection criteria. Our results raise the possibility that host-galaxy dilution contributes to its relatively flat optical continuum, making it a potential analogue of Red Eyes as observed without gravitational lensing. Such an interpretation remains to be tested against intrinsic changes in the accreting source, as also discussed by \citet{Merida+2026}.}

Nevertheless, such systems may still be identified as candidates through careful visual inspection without relying exclusively on strict color or morphology cuts, for example by searching for subtle compact red cores or related substructure. Searches of this kind will therefore be important in future studies. Ultimately, spectroscopic follow-up will be required to determine whether the compact components in visually selected candidates are LRDs rather than ordinary galactic structures or star-forming regions.}

\subsection{Implications for SMBH Formation}
Here we discuss the implications for SMBH formation based on the multiple, off-center, faint LRDs.
To estimate the black hole masses of LRD1 and LRD2, we follow the methodology presented in \cite{Umeda+2025}, assuming 
that the bolometric luminosity of an LRD ($L_\mathrm{bol}$) equals to the luminosity of the blackbody emission. 
This assumption is motivated by the presence of hypothetical gaseous envelopes enshrouding the accreting black hole, whose thermal spectra explain the spectral features of LRDs \citep{Kido+2025, Inayoshi+2025}, and the bolometric correction is supported for some LRDs that have multi-wavelength observations including MIRI photometry \citep{Greene+2025, Ronayne+2025}.
The $L_\mathrm{bol}$ values of LRD1 and LRD2 are shown in Figure \ref{fig:lbol}. For comparison, we also present $L_\mathrm{bol}$ values of LRDs with similar host stellar masses ($M_\star\sim10^{8}\,\mathrm{M_\odot}$) reported in \cite{Umeda+2025}, which are not gravitationally lensed. { LRD1 and LRD2 have $L_\mathrm{bol}= {1.0^{+2.0}_{-0.6}} \times 10^{43}\,\mathrm{erg~s^{-1}}$ and ${2.3^{+6.1}_{-1.5}} \times 10^{43}\,\mathrm{erg~s^{-1}}$, respectively, which are at the faint end or significantly lower than those of previously reported LRDs with similar stellar masses.}

{ We define the Eddington ratio as $\lambda_\mathrm{Edd}\equiv L_\mathrm{bol}/L_\mathrm{Edd}$, where $L_\mathrm{Edd}=4\pi G M_\mathrm{BH}m_\mathrm{p}c/\sigma_\mathrm{T}$ is the electron-scattering Eddington luminosity for fully ionized hydrogen \citep[e.g.,][]{King_Pounds+2015}. Combining these definitions gives}

{
\begin{align}
    M_\mathrm{BH} &= \frac{\sigma_\mathrm{T}}{4 \pi c G m_\mathrm{p}} \cdot \frac{L_\mathrm{bol}}{\lambda_\mathrm{Edd}} \\
    &\simeq 0.81 \times 10^5 \, \mathrm{M_\odot}\,\lambda_\mathrm{Edd}^{-1} \left(\frac{L_\mathrm{bol}}{10^{43}\,\mathrm{erg\,s^{-1}}}\right),
\end{align}
}
where $\sigma_\mathrm{T}$, $c$, $G$, and $m_\mathrm{p}$ are the Thomson scattering cross section, the speed of light, the gravitational constant, and the proton mass, respectively. Following \cite{Umeda+2025}, we assume a fiducial value of 0.5, which is motivated by the near-Eddington accretion phase expected in the BHE model. { This estimate is not an independent virial black-hole mass measurement: it scales linearly with the assumed $L_\mathrm{bol}=L_\mathrm{BB}$ and inversely with the adopted $\lambda_\mathrm{Edd}$. At the fiducial $\lambda_\mathrm{Edd}=0.5$, the inferred black hole masses of LRD1 and LRD2 are $\log (M_\mathrm{BH}/\mathrm{M_\odot})={5.19^{+0.49}_{-0.41}}$ and ${5.57^{+0.56}_{-0.44}}$, respectively, which put them in the intermediate-mass black hole (IMBH) regime.} If faint, off-center, multiple LRDs are commonly present in typical SFGs, several possibilities may be considered. one potential scenario is that multiple IMBHs form at off-nuclear locations, possibly within massive star clusters \citep[e.g.,][]{PortegiesZwart+2002, PortegiesZwart+2004, Sakurai+2017, Fujii+2024}, and subsequently migrate toward the galaxy center by dynamical friction, eventually contributing to the growth of a central SMBH via mergers \citep[e.g.,][]{Dekel+2025}. Following \cite{Binney_Tremaine_2008} and \cite{Ubler+2025}, we estimate the dynamical friction timescale, $t_\mathrm{df}$, for the two LRDs to migrate to the galaxy center using the following equation:

\begin{figure}[!t]
    \centering
    \includegraphics[width=1\linewidth]{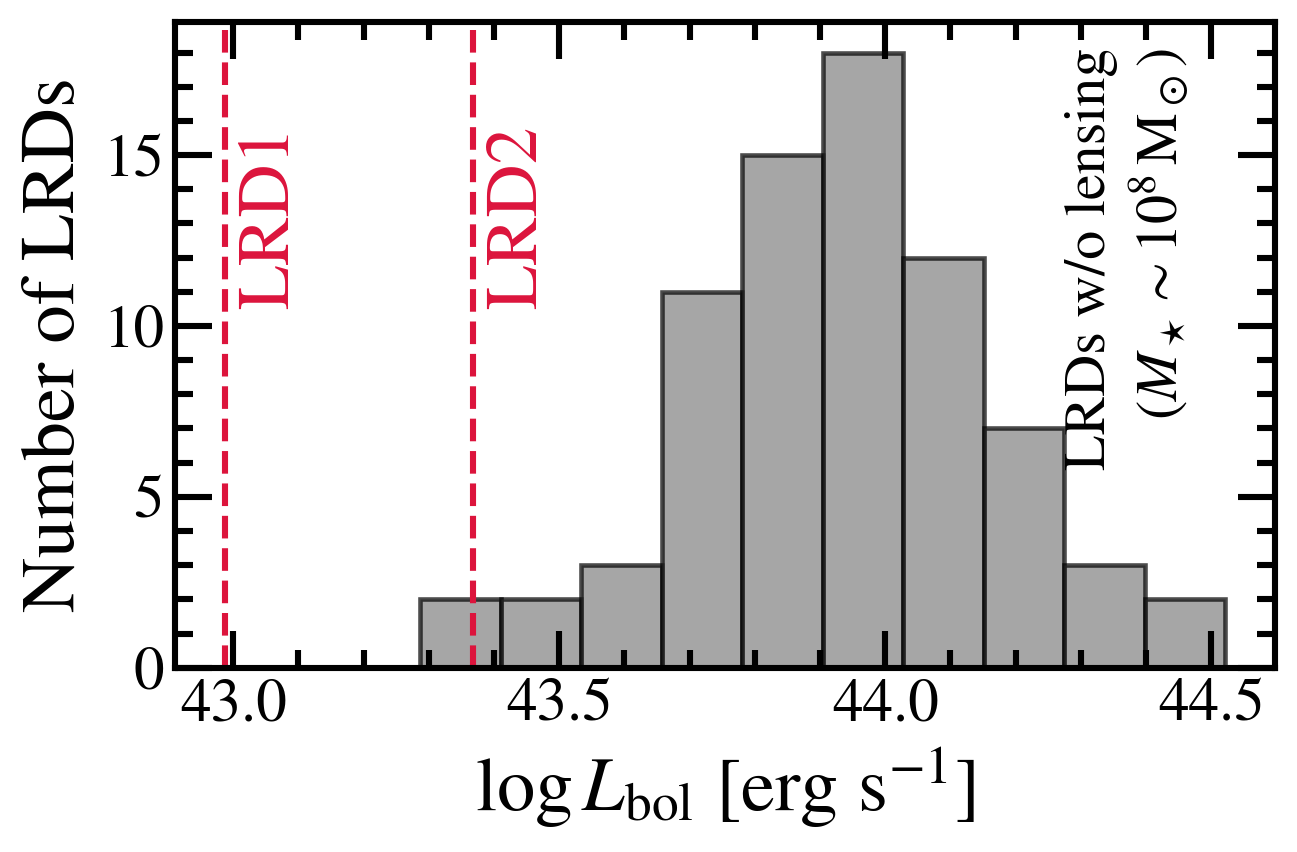}
    \caption{Bolometric luminosity distribution of LRDs. The red dashed lines indicate LRD1 and LRD2. The gray histogram shows the LRDs presented in \cite{Umeda+2025}, who estimate $L_\mathrm{bol}$ of the LRD sample presented in \cite{Akins+2025} based on the black hole envelope model \citep{Kido+2025, Inayoshi+2025}.}
    \label{fig:lbol}
\end{figure}

\begin{align}
    t_\mathrm{df} = \frac{1.65}{\ln\Lambda} \frac{R^2 \sigma_\star}{GM_\mathrm{BH}},
\end{align}
where $R$ is the distance between the LRD and the galactic center ($\sim{900}~\mathrm{pc}$), and $\sigma_\star$ is the stellar velocity dispersion. { We assume $\sigma_\star=20~\mathrm{km~s^{-1}}$ based on the $M_\mathrm{BH}$--$M_\star$ and $M_\mathrm{BH}$--$\sigma_\star$ relations from \cite{Greene+2020} at a stellar mass of $M_\star\sim10^8\,\mathrm{M_\odot}$. We use $M_\star=2.7\times10^8\,\mathrm{M_\odot}$ to evaluate the Coulomb logarithm, $\ln\Lambda$, with $\Lambda \approx M_\star/M_\mathrm{BH}$. For the fiducial $\lambda_\mathrm{Edd}=0.5$ masses, we obtain $t_\mathrm{df}\simeq4.7\,\mathrm{Gyr}$ for LRD1 and $3.0\,\mathrm{Gyr}$ for LRD2. Starting from $z=7.4$, these times correspond to approximate merger redshifts of $z\simeq1.1$ and $1.8$, respectively.} The off-center faint LRDs can thus migrate to the galactic center within the age of the Universe, possibly contributing to SMBH growth via mergers.

{ The multiple, off-nuclear configuration of Red Eyes may be qualitatively consistent with a dense-star-cluster origin of faint LRDs. Runaway stellar collisions in hierarchically assembling clusters can produce very or supermassive stars and, subsequently, IMBH seeds or quasi-stars, while neighboring sub-clusters remain visible as blue star-forming clumps \citep{Chisholm+2026,Rantala2026}. A related possibility is that collisions between stellar-mass black holes and massive stars in dense stellar environments form quasi-hydrostatic ``black hole stars,'' providing another pathway to rapid black-hole growth that may be relevant to LRDs \citep{Shi+2026}. 
Cosmological simulations predict the nearly simultaneous formation of multiple heavy seeds followed by their inward migration and merger \citep{Chon+2026}. Moreover, ultra-deep SPURS spectroscopy of GN-z11 provides observational evidence that compact, low-metallicity starbursts can host very massive stars and may constitute efficient IMBH-seeding sites (\citealp{Chen+2026}; see also \citealp{Nakane_Ouchi_2026}). Red Eyes may therefore represent a lower-mass, spatially resolved candidate for black holes forming in compact star clusters. However, the required cluster densities and the predicted stellar-wind, abundance, and accretion signatures are not currently constrained, so this interpretation remains speculative.} {A somewhat more exotic variant of the multiple seed formation scenario could involve massive primordial black holes that facilitates the formation of a nearby direct-collapse black hole \citep[e.g.,][]{Zhang2025_PBH}.}

{Another possibility is that the LRD pair is interacting or merging with the blue SFG, reported around a substantial fraction of LRDs \citep[e.g.,][]{Golubchik+2025,Baggen+2025,baggen_connecting_2026, Barger+2026}, potentially enhancing the gas supply to the putative accreting black holes. \citet{Merida+2026} suggest that interactions with neighboring galaxies may be linked to the emergence or fading of an LRD phase. \citet{Barger+2026} likewise report frequent close companions around LRDs and broad-line AGNs, often comprising a red source and a blue companion, and propose merger-driven accretion as a possible explanation. Spatially resolved spectroscopy is needed to constrain the relative kinematics and gas distribution and assess whether interactions contribute to fueling the LRDs.}

Alternatively, the low $L_\mathrm{bol}$ values may be attributed to low $\lambda_\mathrm{Edd}$, rather than intrinsically low black hole masses, implying an inactive or weakly accreting phase. {For example, if we assume $\lambda_\mathrm{Edd}=0.1$, we obtain $M_\mathrm{BH}={(0.77\text{--}1.8)\times10^6}\,\mathrm{M_\odot}$, which is comparable to those inferred for the typical LRDs \citep[e.g.,][]{Harikane+2023, Matthee+2024, Greene+2024, Kocevski+2025, Rusakov+2026}.} \cite{Inayoshi+2025} argue that the LRD is sustained by near-Eddington accretion onto a neutral gas envelope surrounding the black hole, and that the LRD phase terminates once the accretion rate drops well below the Eddington limit. In this context, Red Eyes may represent a system observed at the very end of the LRD phase, providing a unique opportunity to investigate the evolutionary connection between LRDs and normal AGNs or QSOs.

These interpretations remain speculative because the current broadband photometry lacks key spectroscopic evidence, such as broad emission lines, needed to confirm accreting black holes and constrain their masses. Further spectroscopic follow-up is therefore essential to test the nature of Red Eyes and the SMBH-formation scenarios discussed above.




\section{Summary}\label{sec:summary}
In this work, we present the detection of a very close pair of LRDs found in a strongly lensed galaxy, which are embedded in the typical SFGs at $z\sim7$. The main findings in this work are summarized below:

\begin{itemize}
    \item We identify a strongly lensed arc ($\mu \sim${11--18}) at $z \sim 7$ that contains two point-like objects satisfying the LRD color selection criteria. The critical curve at $z \sim 7$ does not pass between the two LRD images, indicating that they are not multiple images of a single source, but instead correspond to two distinct objects on the source plane. In contrast, the extended arc exhibits a blue SED, implying that it is a star-forming galaxy associated with the LRDs. Source-plane reconstruction reveals that the two LRDs are separated by only $\sim{150}\,\mathrm{pc}$ and are offset from the galaxy center by $\sim{900}\,\mathrm{pc}$.

    \item The intrinsic SEDs of the two LRDs are much fainter than that of the associated galaxy, indicating that the flux from the LRDs would be {diluted} by that of the host galaxy
    in the absence of gravitational lensing. Mock NIRCam observations further show that such a system would resemble typical star-forming galaxies at similar redshifts without gravitational lensing. These analyses suggest that, without magnification, many of these LRDs may be unnoticed inside normal SFGs.

\end{itemize}

\clearpage
\appendix
\section{{SED-fitting posterior distributions}}\label{app:sed-posteriors}

\begin{figure*}[p]
    \centering
    \includegraphics[width=\textwidth]{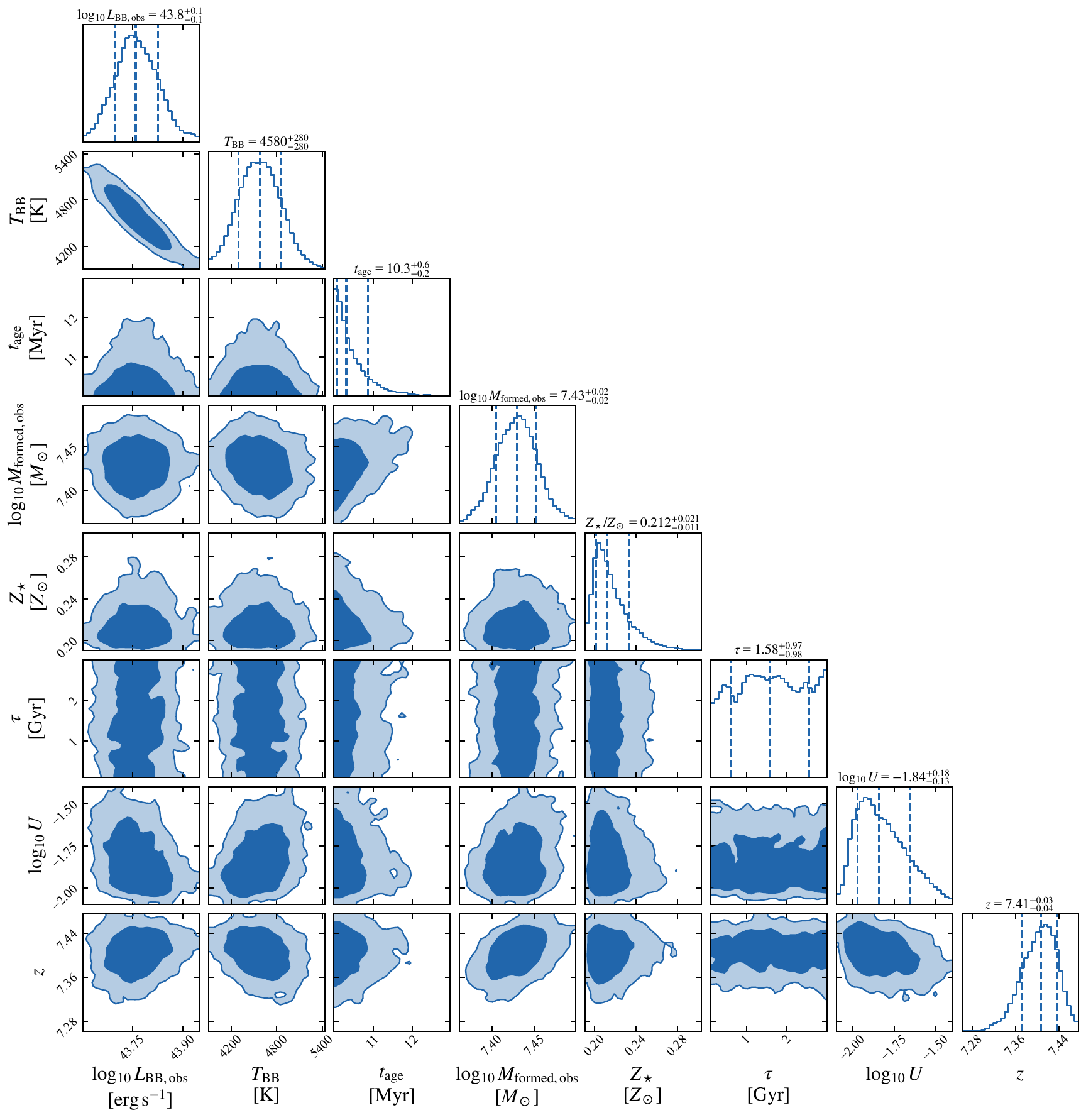}
    \caption{ SED-fitting posterior distribution for LRD-SW1.}
    \label{fig:corner-lrd-sw1}
\end{figure*}

\begin{figure*}[p]
    \centering
    \includegraphics[width=\textwidth]{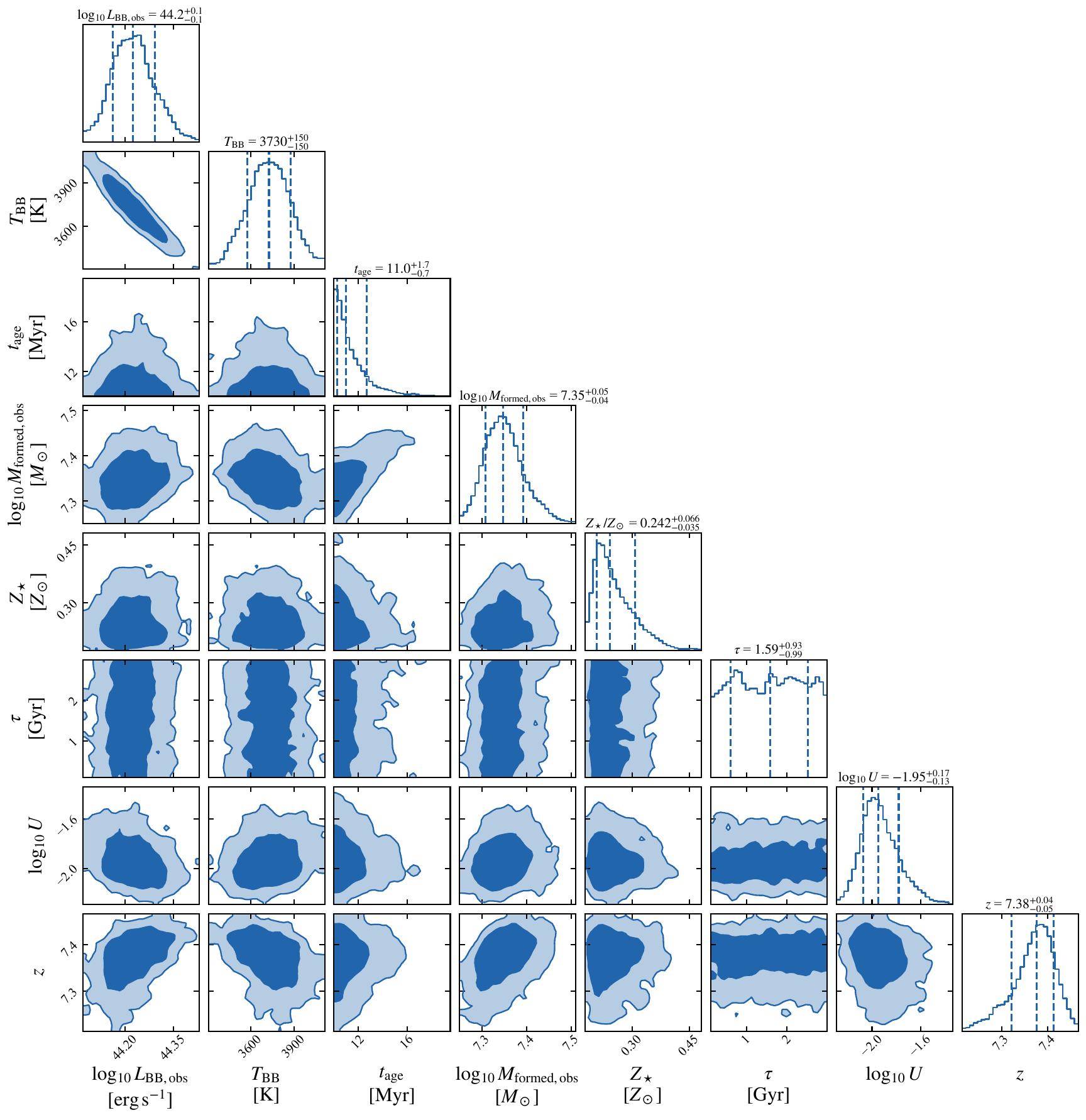}
    \caption{ SED-fitting posterior distribution for LRD-SW2.}
    \label{fig:corner-lrd-sw2}
\end{figure*}

\begin{figure*}[p]
    \centering
    \includegraphics[width=\textwidth]{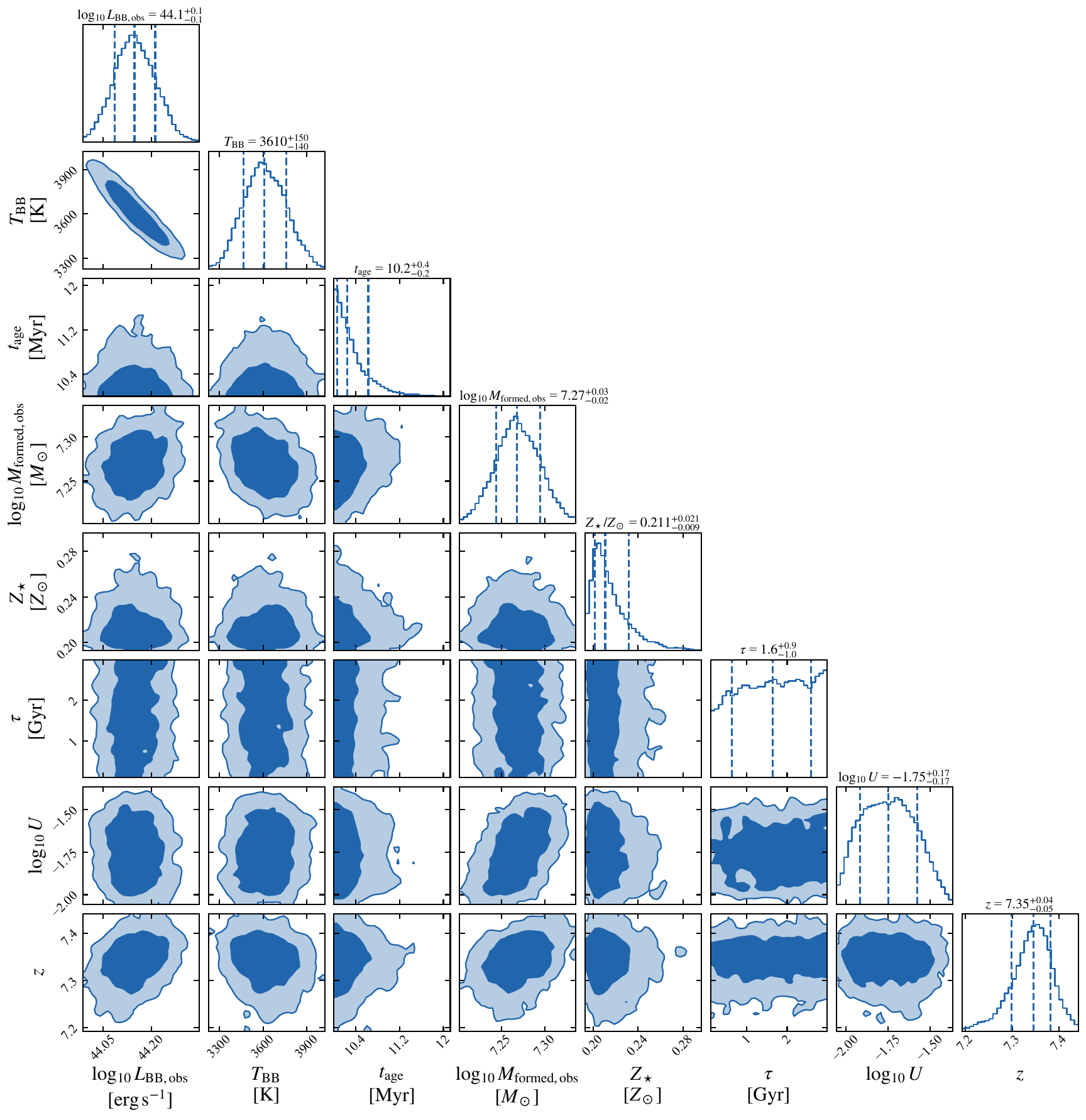}
    \caption{ SED-fitting posterior distribution for LRD-NW.}
    \label{fig:corner-lrd-nw}
\end{figure*}

\begin{figure*}[p]
    \centering
    \includegraphics[width=\textwidth]{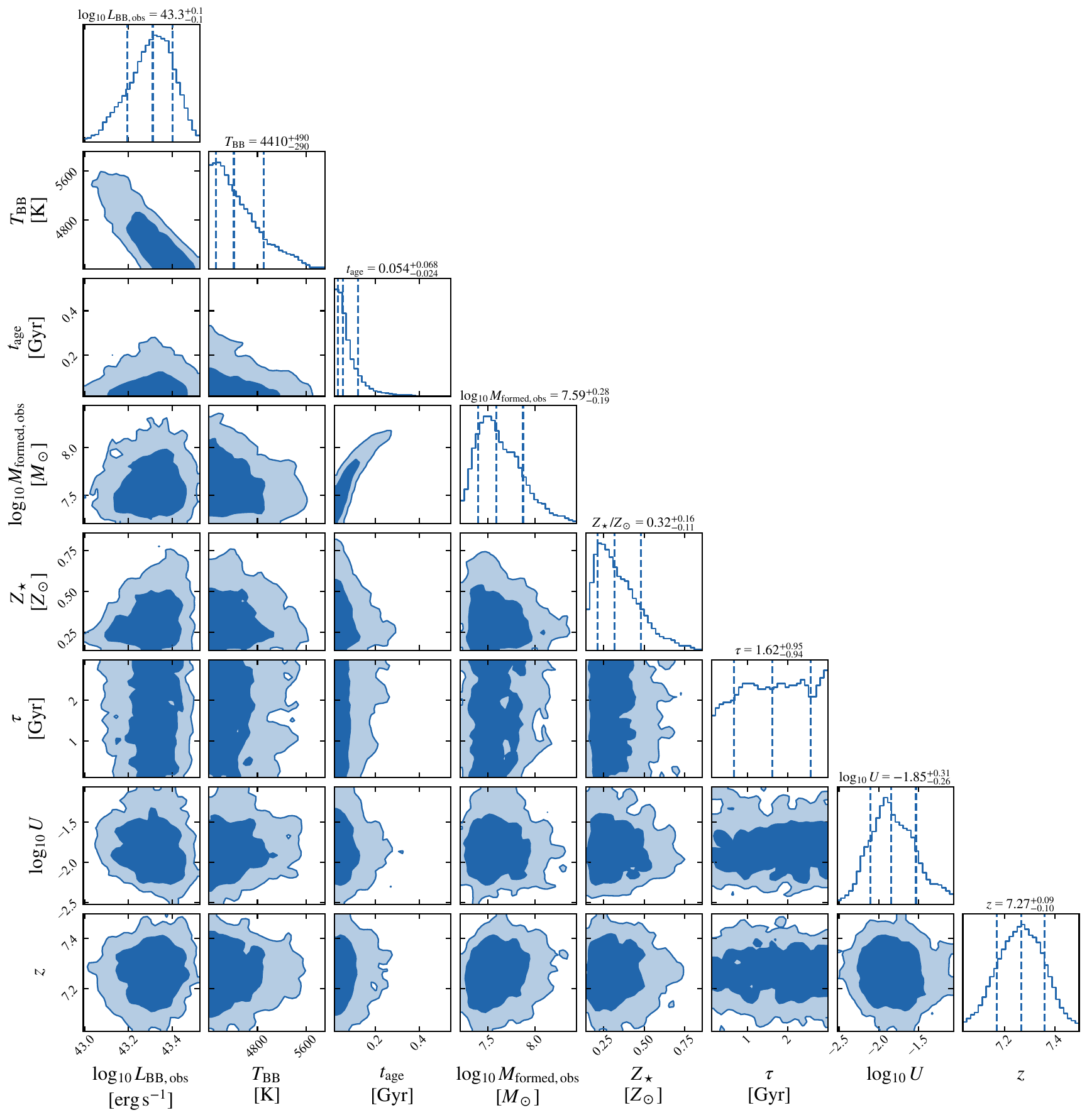}
    \caption{ SED-fitting posterior distribution for SFG-NE1.}
    \label{fig:corner-sfg-ne1}
\end{figure*}

\begin{figure*}[p]
    \centering
    \includegraphics[width=\textwidth]{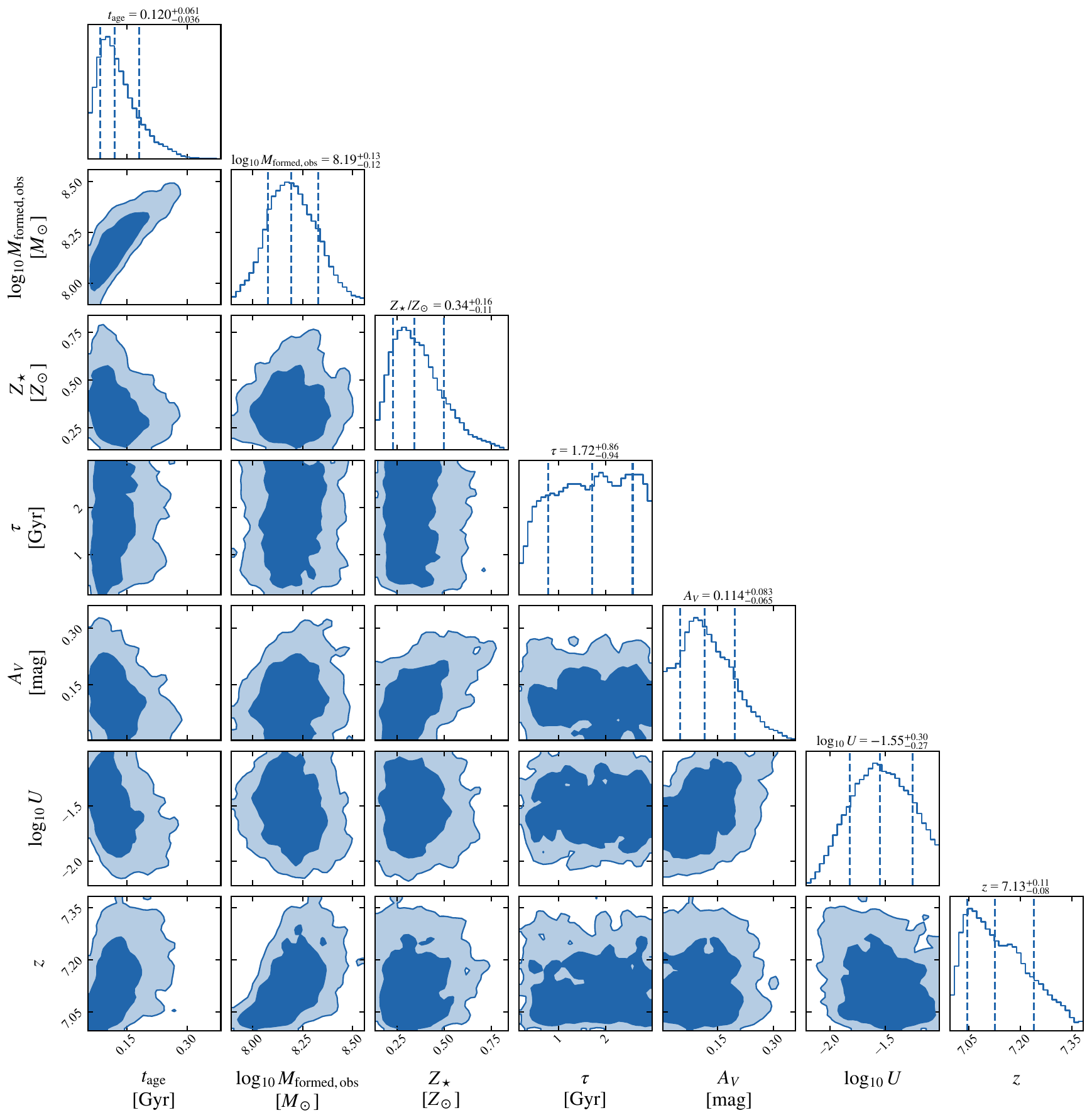}
    \caption{ SED-fitting posterior distribution for SFG-NE2.}
    \label{fig:corner-sfg-ne2}
\end{figure*}

\begin{figure*}[p]
    \centering
    \includegraphics[width=\textwidth]{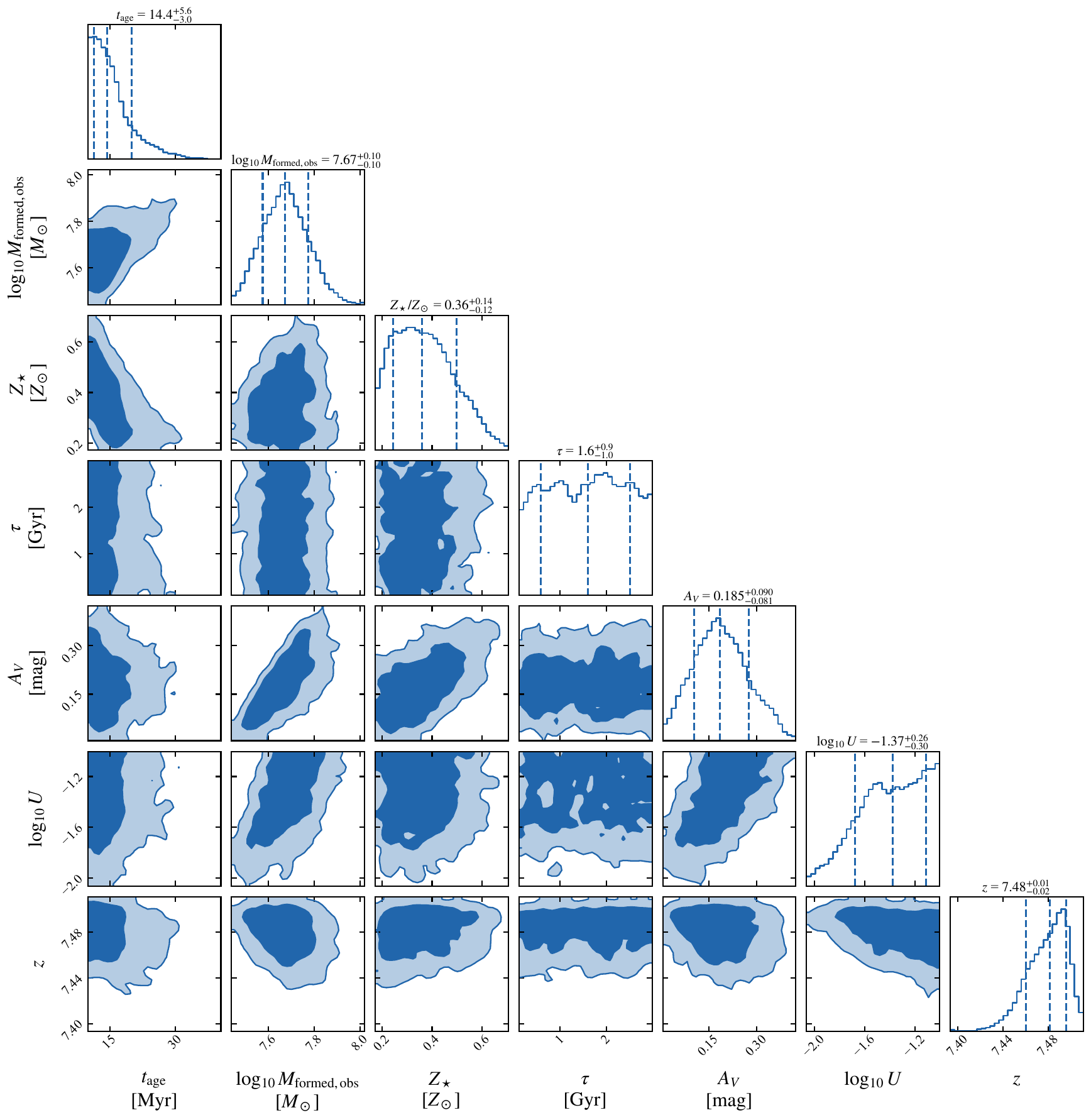}
    \caption{ SED-fitting posterior distribution for SFG-SW1.}
    \label{fig:corner-sfg-sw1}
\end{figure*}

\begin{figure*}[p]
    \centering
    \includegraphics[width=\textwidth]{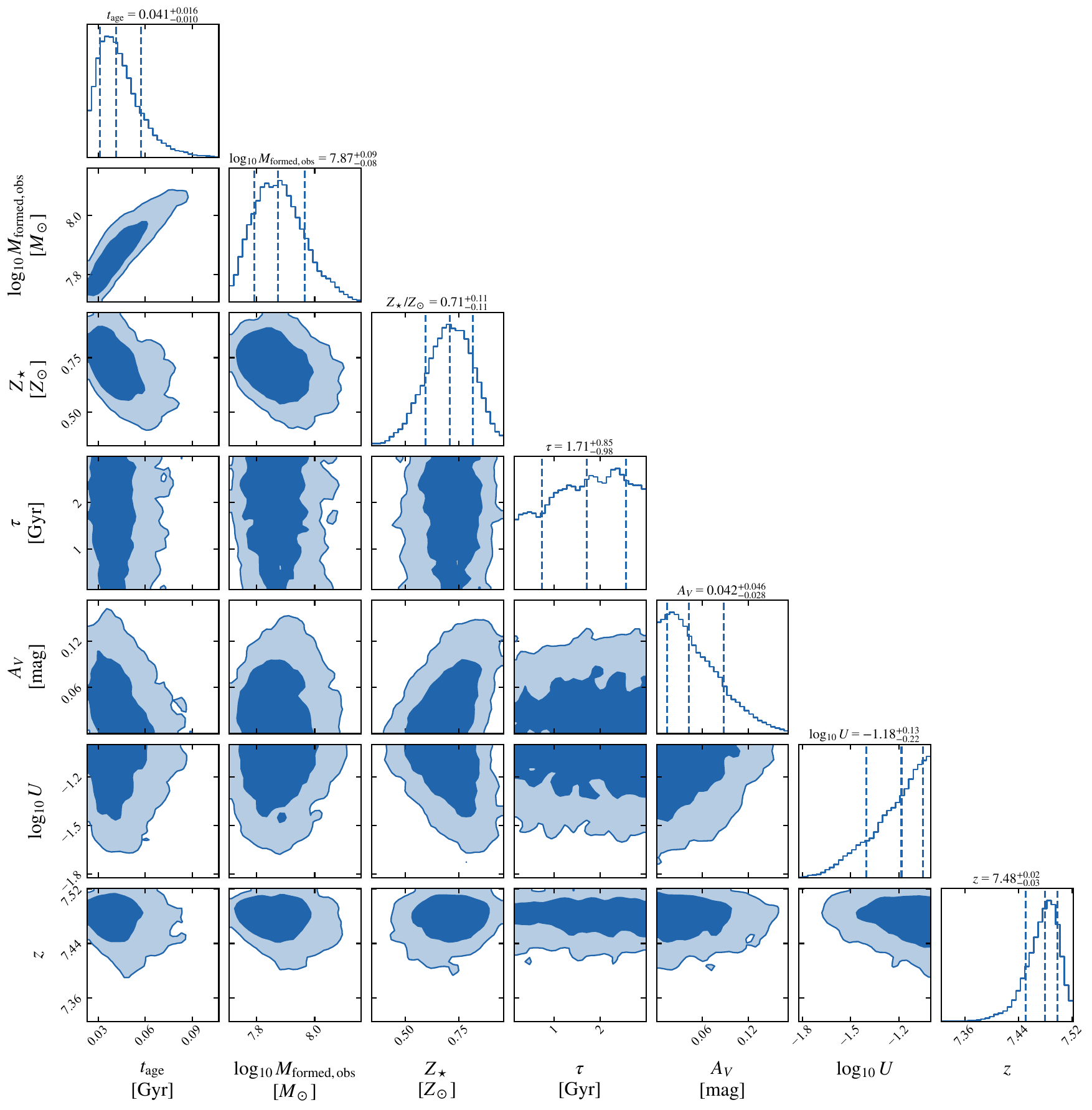}
    \caption{ SED-fitting posterior distribution for SFG-SW2.}
    \label{fig:corner-sfg-sw2}
\end{figure*}

\begin{figure*}[p]
    \centering
    \includegraphics[width=\textwidth]{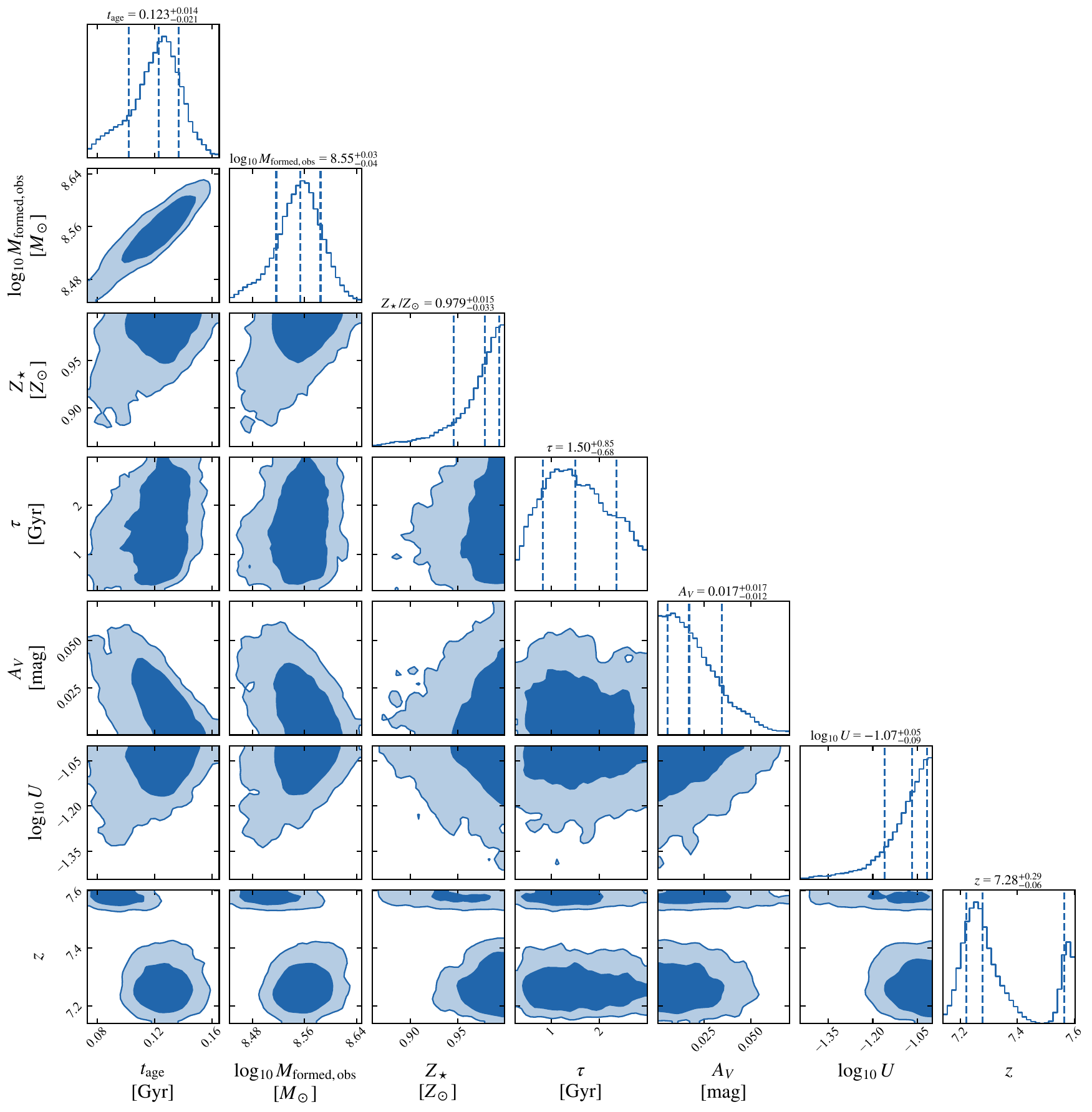}
    \caption{ SED-fitting posterior distribution for SFG-SW3.}
    \label{fig:corner-sfg-sw3}
\end{figure*}

\begin{figure*}[p]
    \centering
    \includegraphics[width=\textwidth]{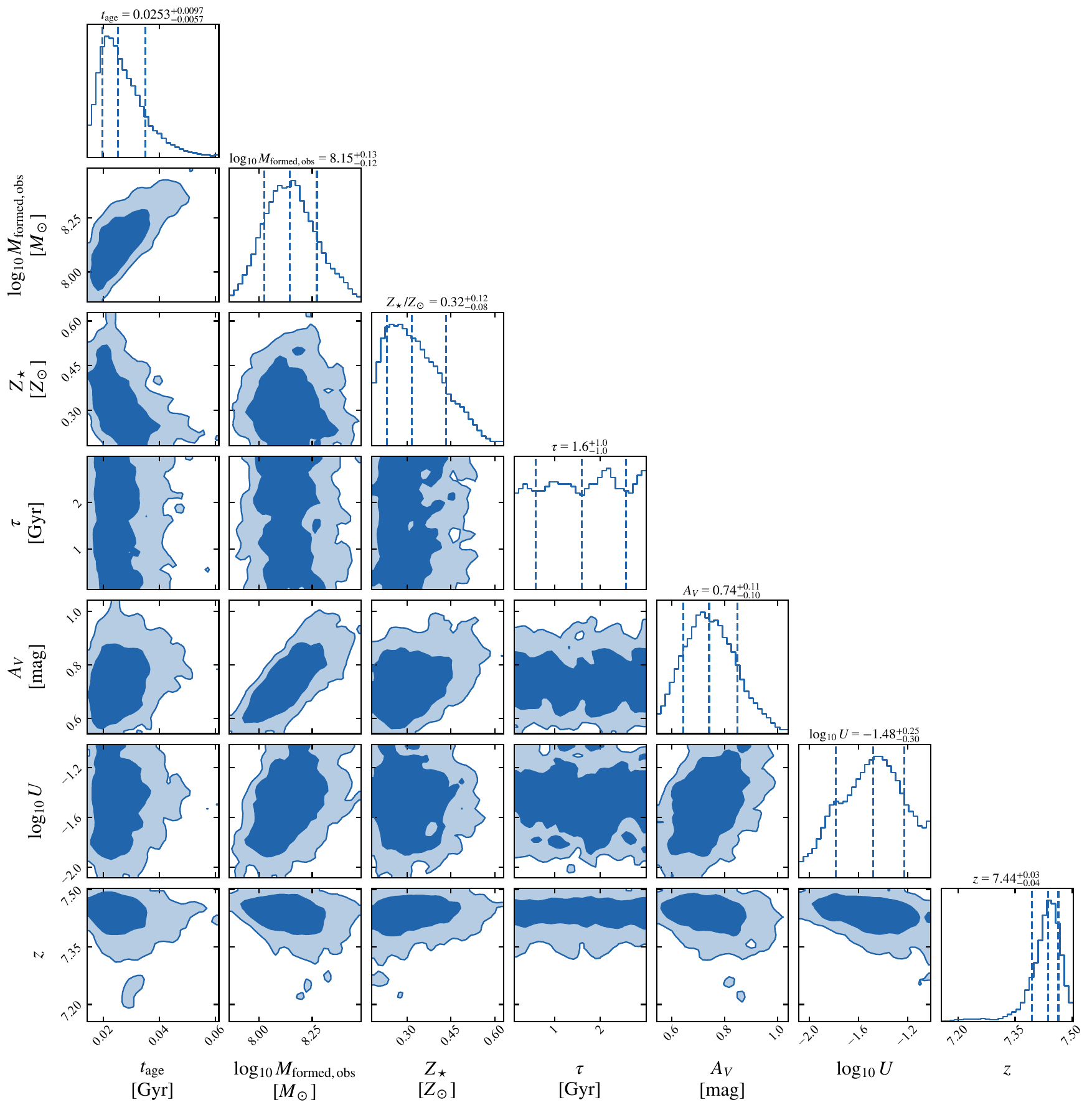}
    \caption{ SED-fitting posterior distribution for SFG-SW4.}
    \label{fig:corner-sfg-sw4}
\end{figure*}

\begin{figure*}[p]
    \centering
    \includegraphics[width=\textwidth]{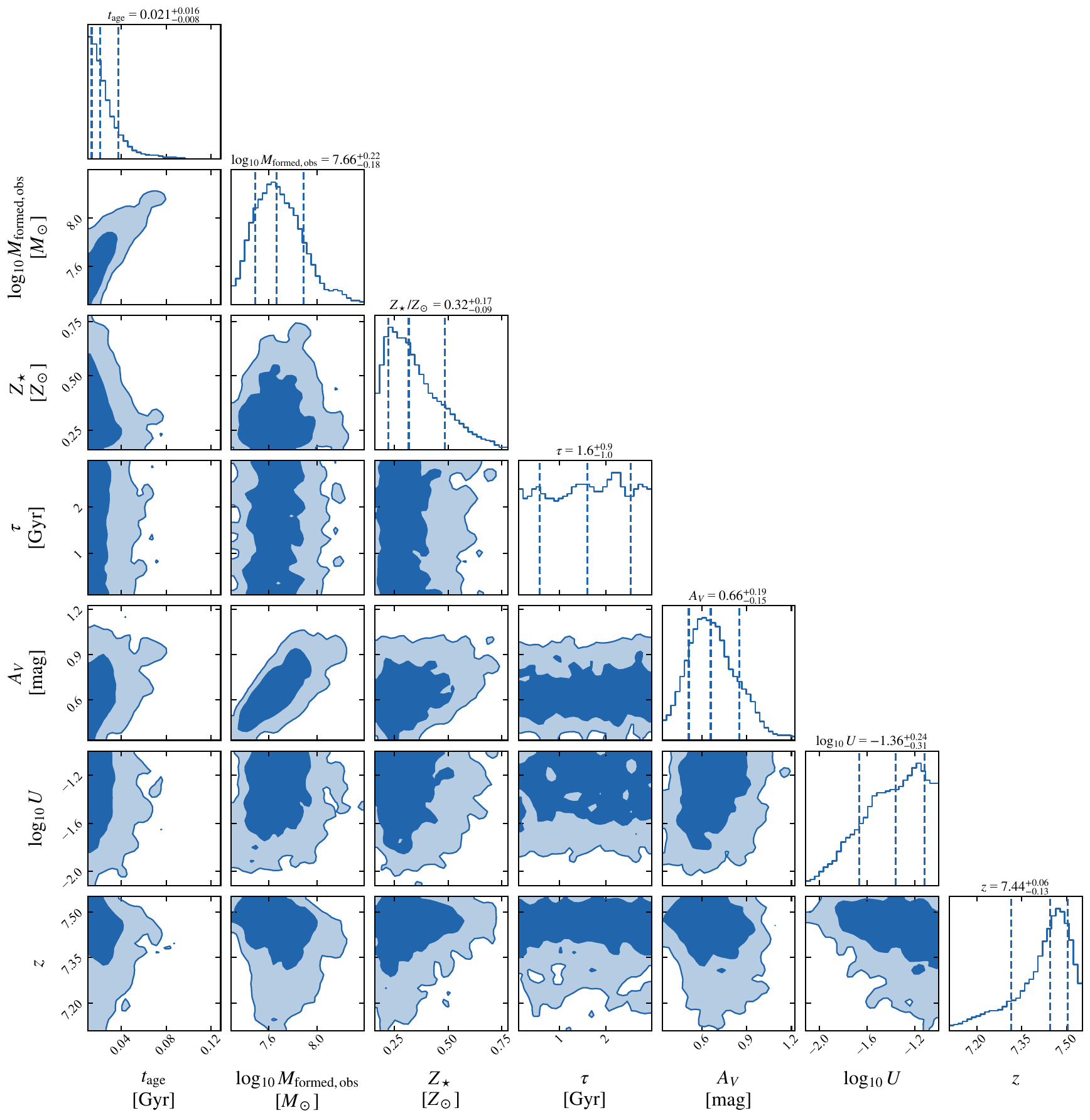}
    \caption{ SED-fitting posterior distribution for SFG-SW5.}
    \label{fig:corner-sfg-sw5}
\end{figure*}

\begin{figure*}[p]
    \centering
    \includegraphics[width=\textwidth]{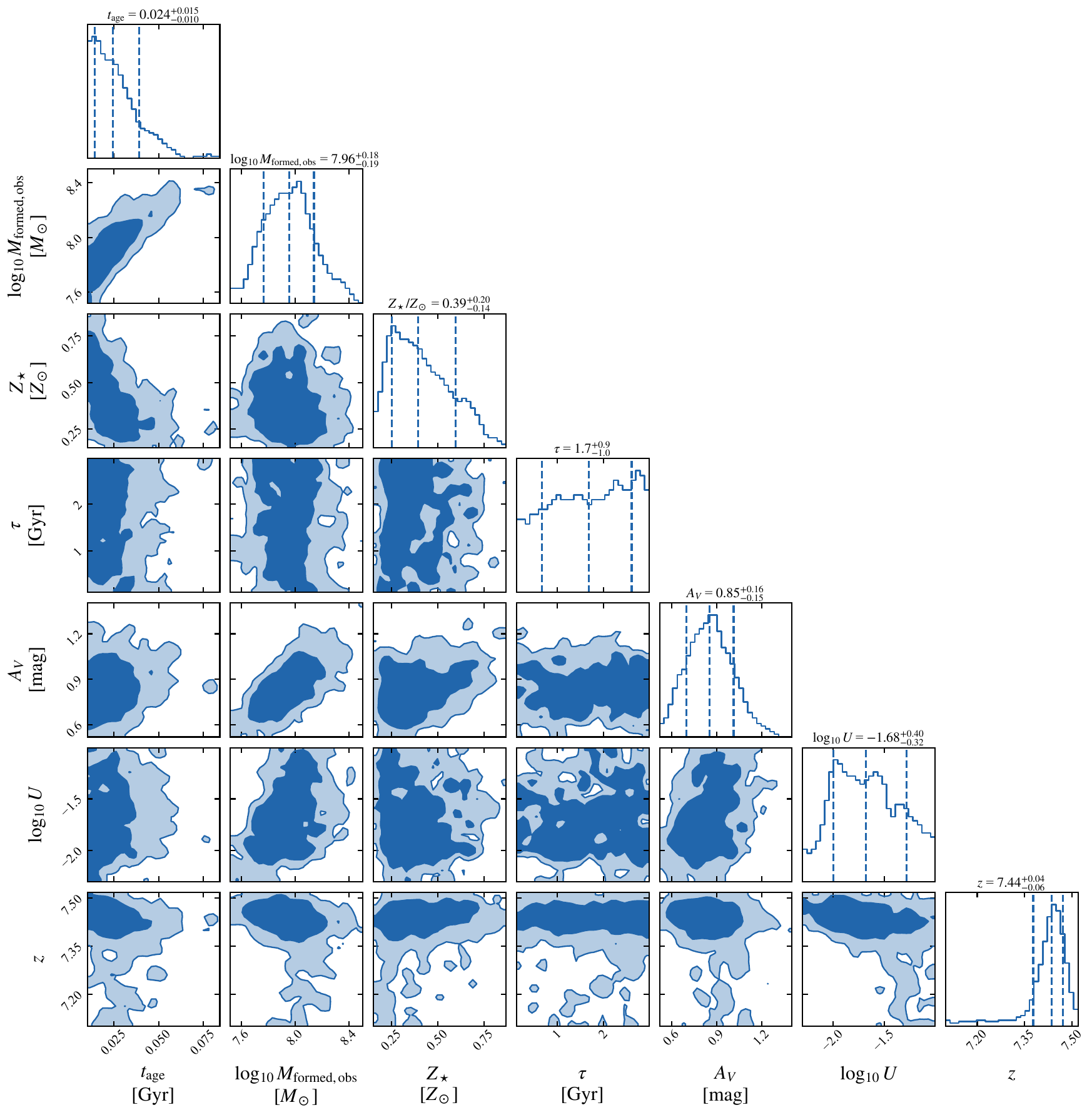}
    \caption{ SED-fitting posterior distribution for SFG-SW6.}
    \label{fig:corner-sfg-sw6}
\end{figure*}

\begin{figure*}[p]
    \centering
    \includegraphics[width=\textwidth]{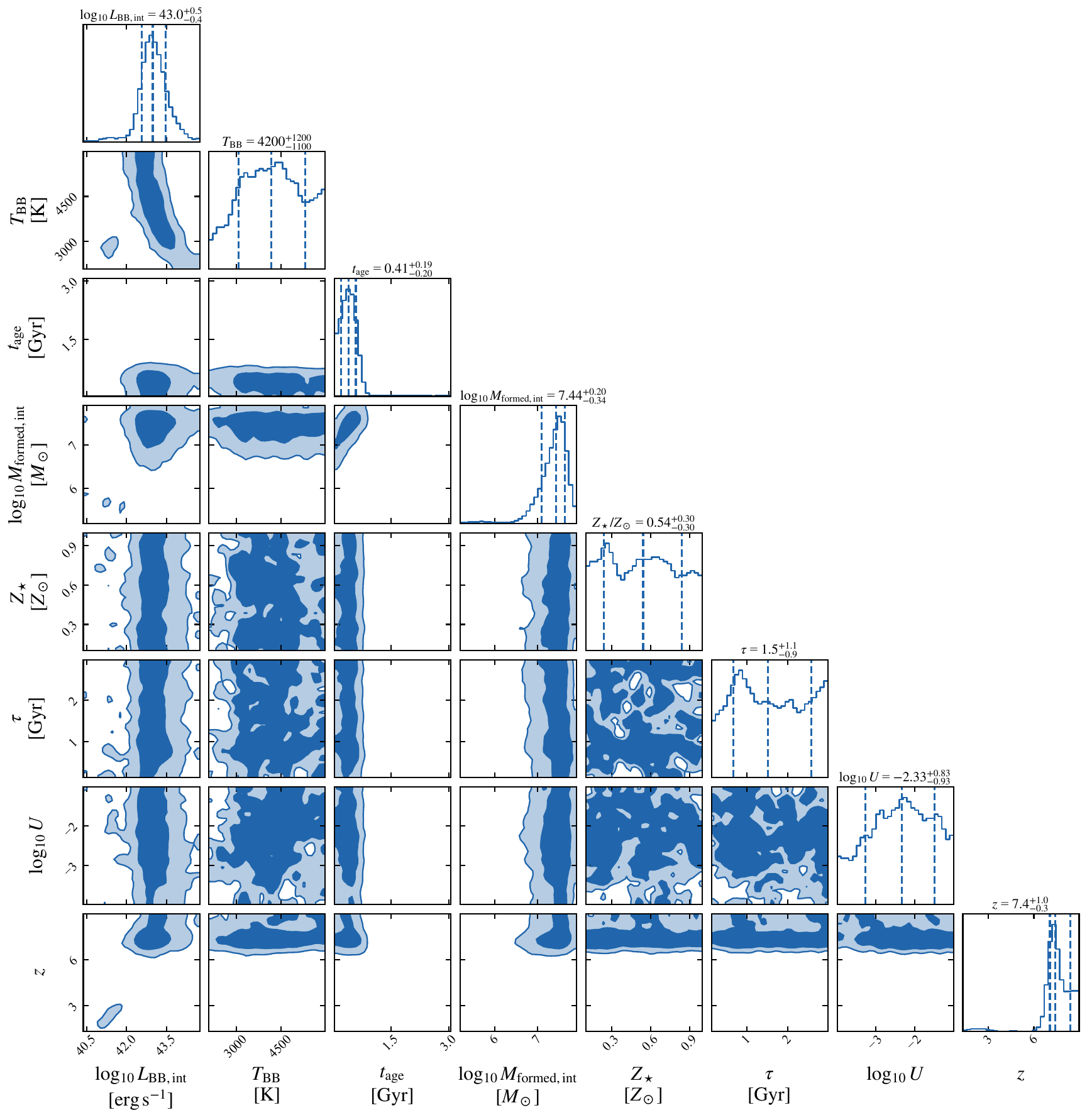}
    \caption{ SED-fitting posterior distribution for intrinsic LRD1 using the galaxy-plus-blackbody model shown in Figure~\ref{fig:sed-int}.}
    \label{fig:corner-figure8-lrd1}
\end{figure*}

\begin{figure*}[p]
    \centering
    \includegraphics[width=\textwidth]{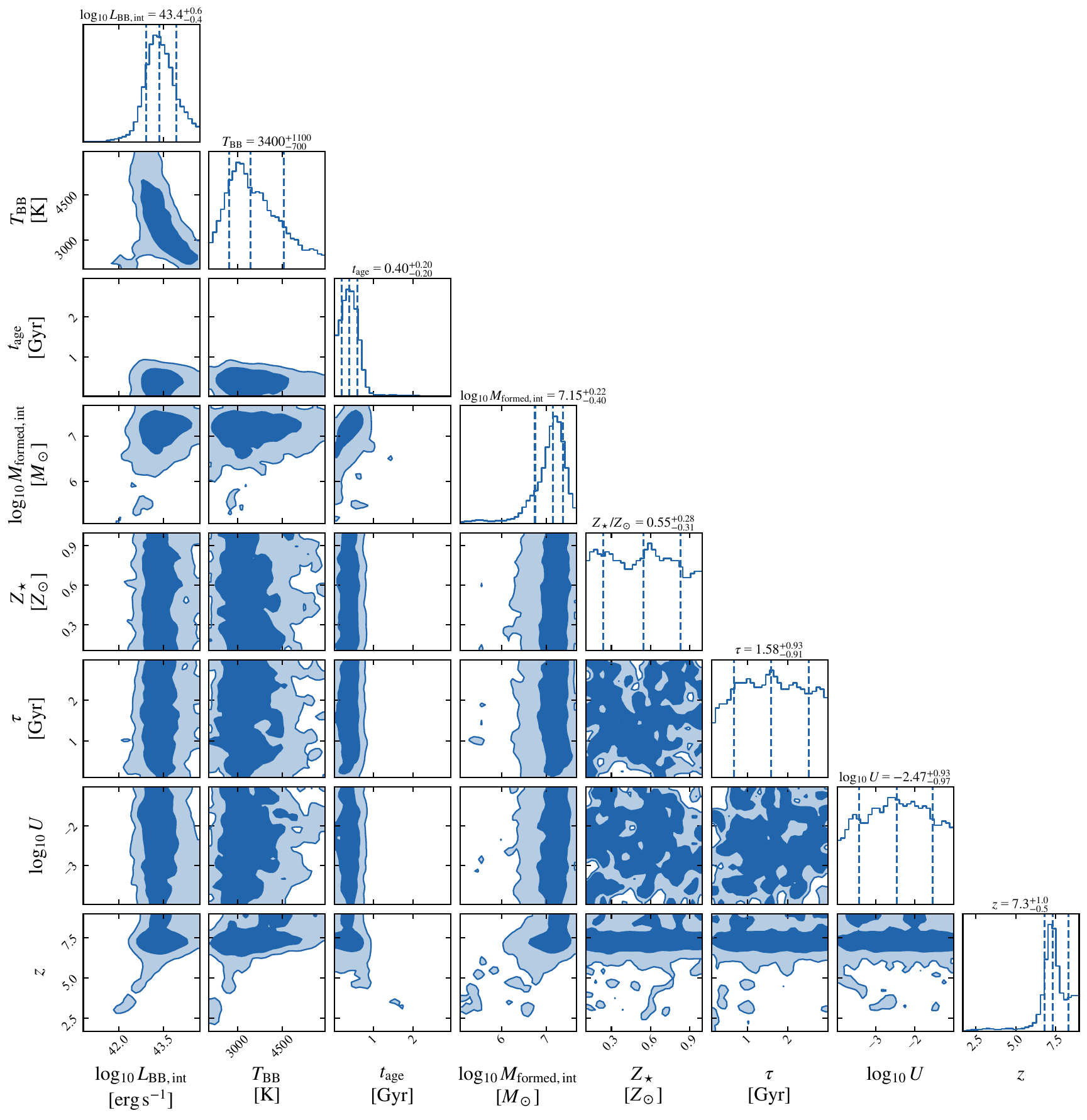}
    \caption{ SED-fitting posterior distribution for intrinsic LRD2 using the galaxy-plus-blackbody model shown in Figure~\ref{fig:sed-int}.}
    \label{fig:corner-figure8-lrd2}
\end{figure*}

\begin{figure*}[p]
    \centering
    \includegraphics[width=\textwidth]{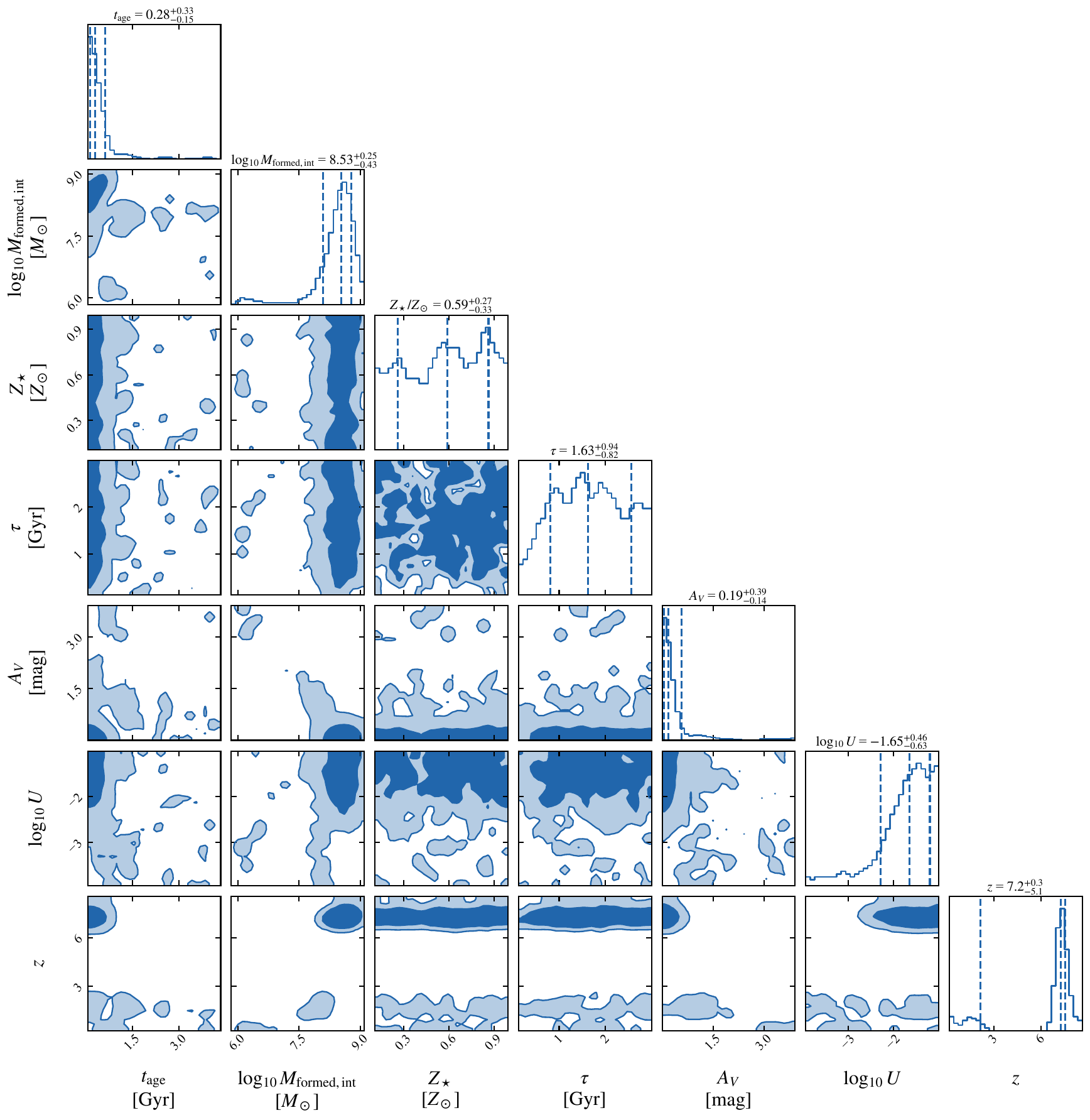}
    \caption{ SED-fitting posterior distribution for the intrinsic SFG total using the galaxy-only model shown in Figure~\ref{fig:sed-int}.}
    \label{fig:corner-figure8-sfg}
\end{figure*}

\begin{figure*}[p]
    \centering
    \includegraphics[width=\textwidth]{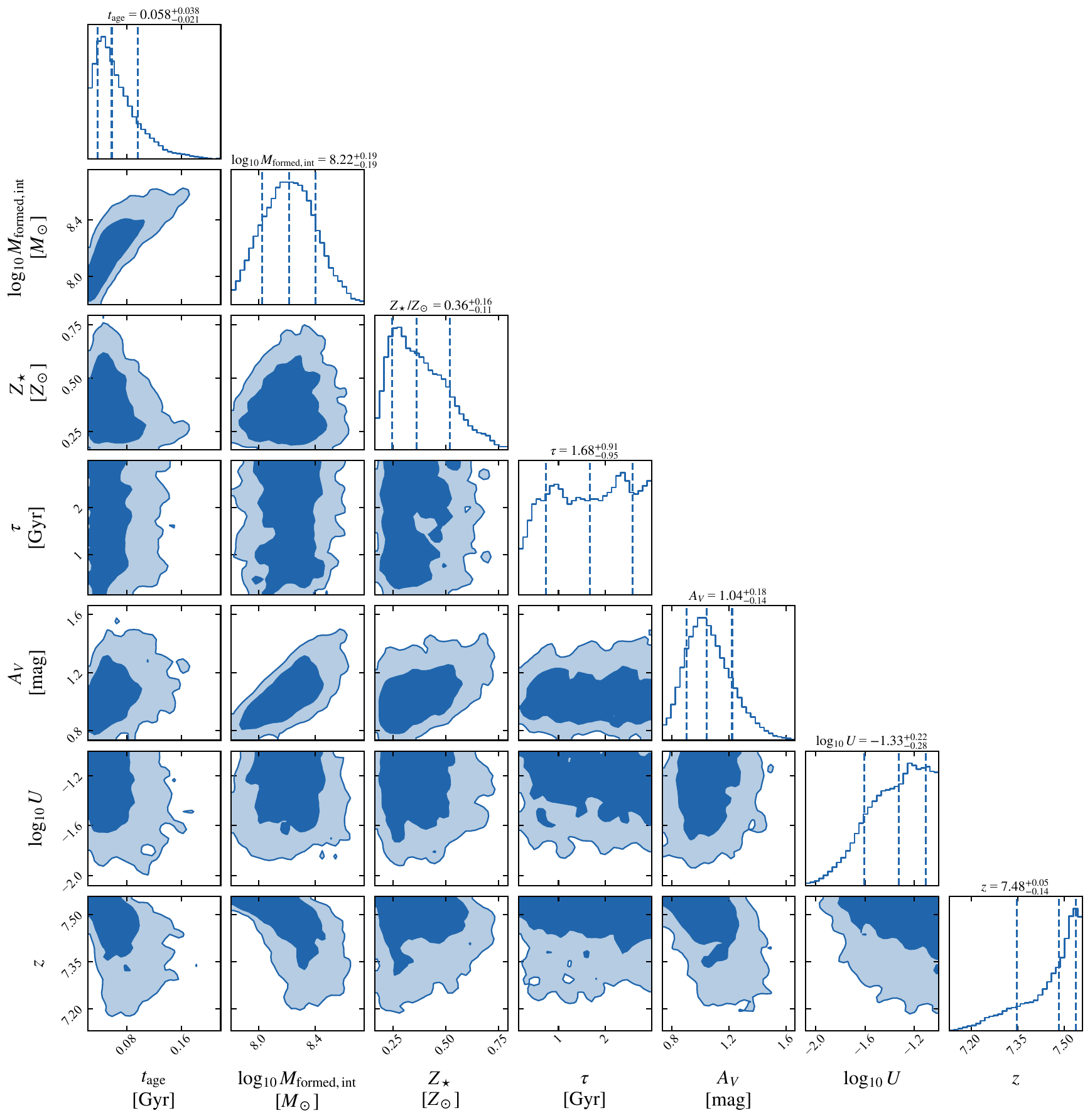}
    \caption{ SED-fitting posterior distribution for the ``unlensed red'' $0.3''$-diameter aperture using the galaxy-only model shown in Figure~\ref{fig:sed-unlensed}.}
    \label{fig:corner-figure11-ap030-gal}
\end{figure*}

\begin{figure*}[p]
    \centering
    \includegraphics[width=\textwidth]{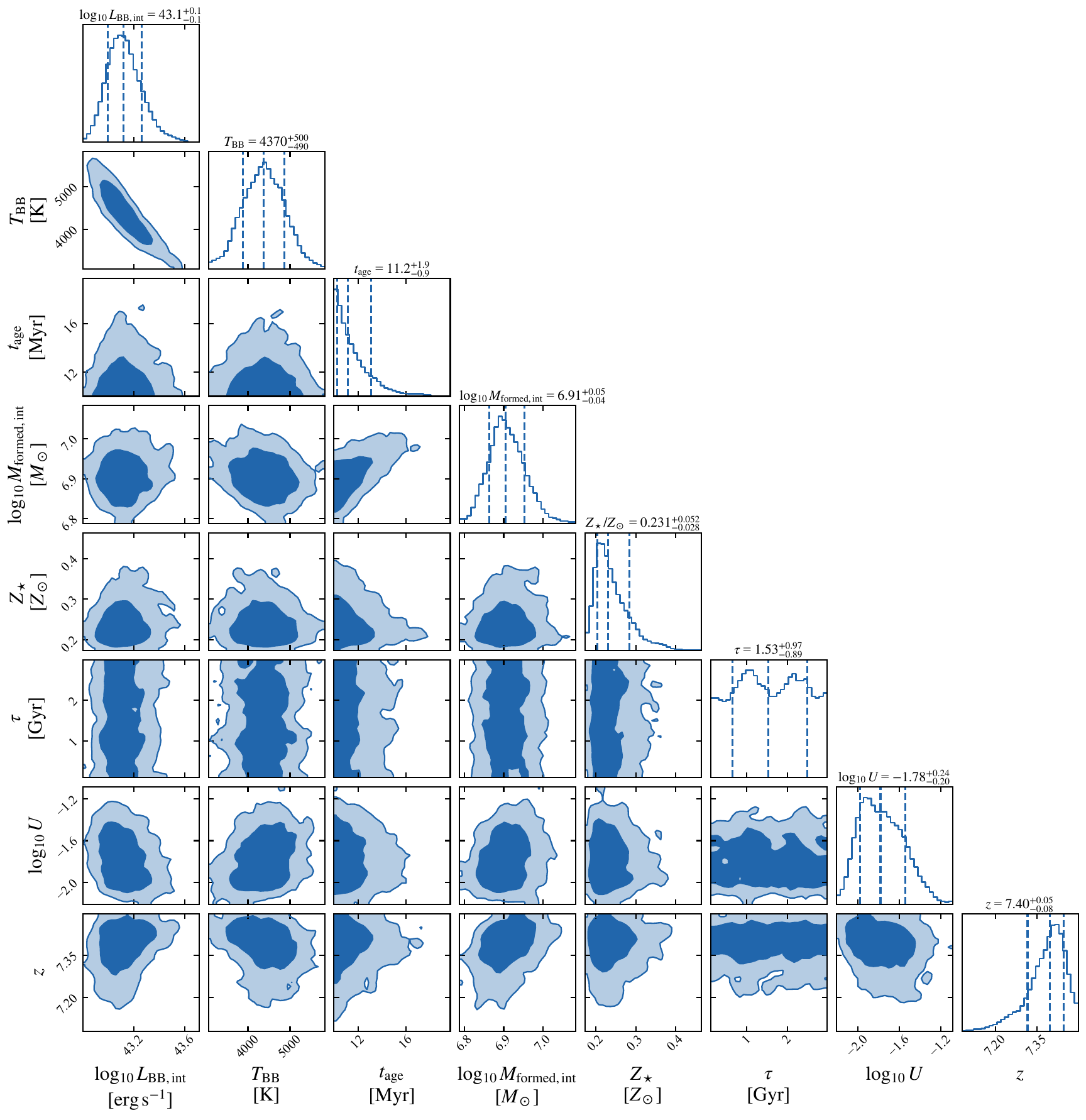}
    \caption{ SED-fitting posterior distribution for the ``unlensed red'' $0.3''$-diameter aperture using the galaxy-plus-blackbody model shown in Figure~\ref{fig:sed-unlensed}.}
    \label{fig:corner-figure11-ap030-bb}
\end{figure*}

\begin{figure*}[p]
    \centering
    \includegraphics[width=\textwidth]{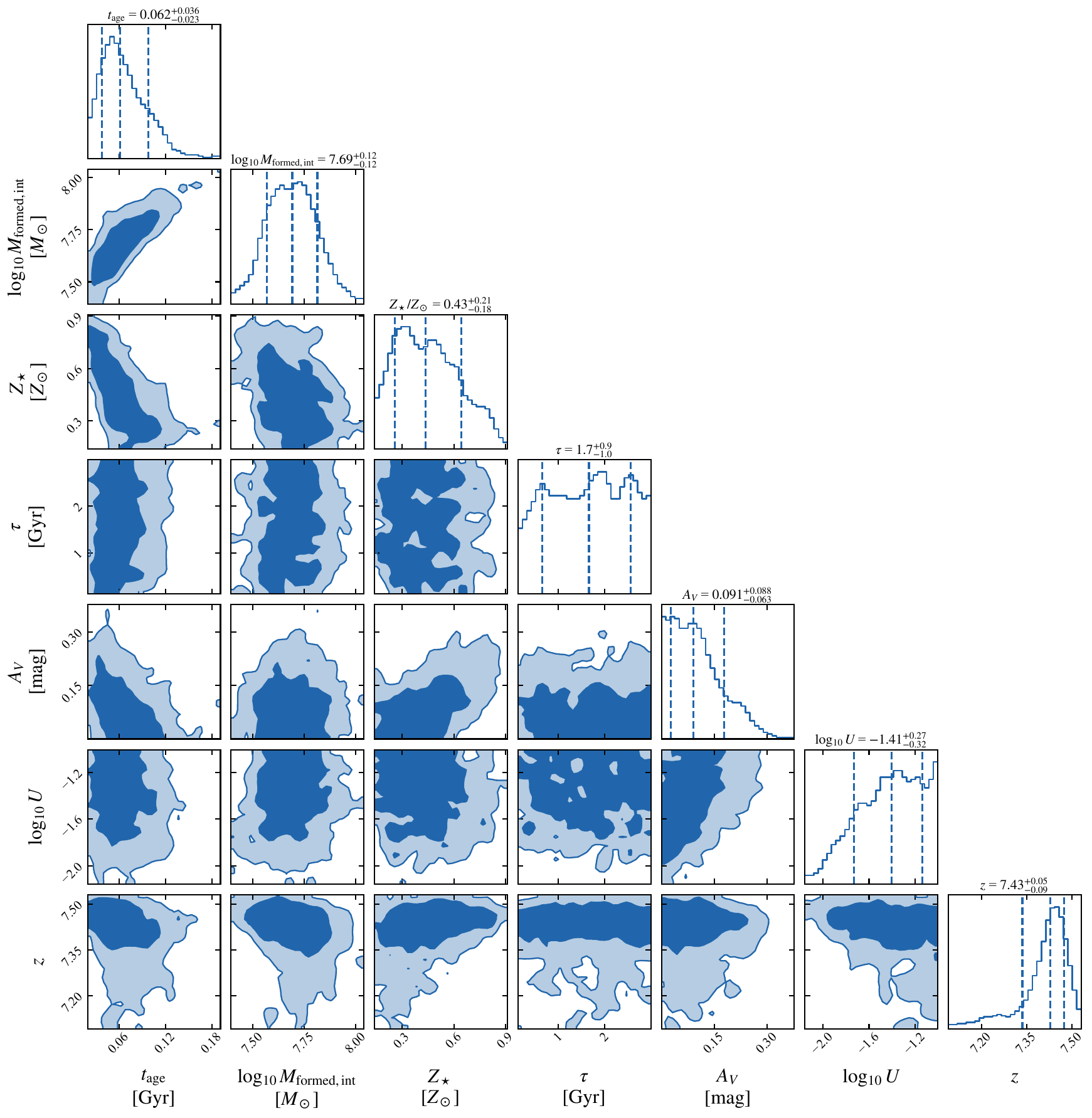}
    \caption{ SED-fitting posterior distribution for the ``unlensed blue'' $0.3''$-diameter aperture using the galaxy-only model shown in Figure~\ref{fig:sed-unlensed}.}
    \label{fig:corner-figure11-blue-gal}
\end{figure*}

\begin{figure*}[p]
    \centering
    \includegraphics[width=\textwidth]{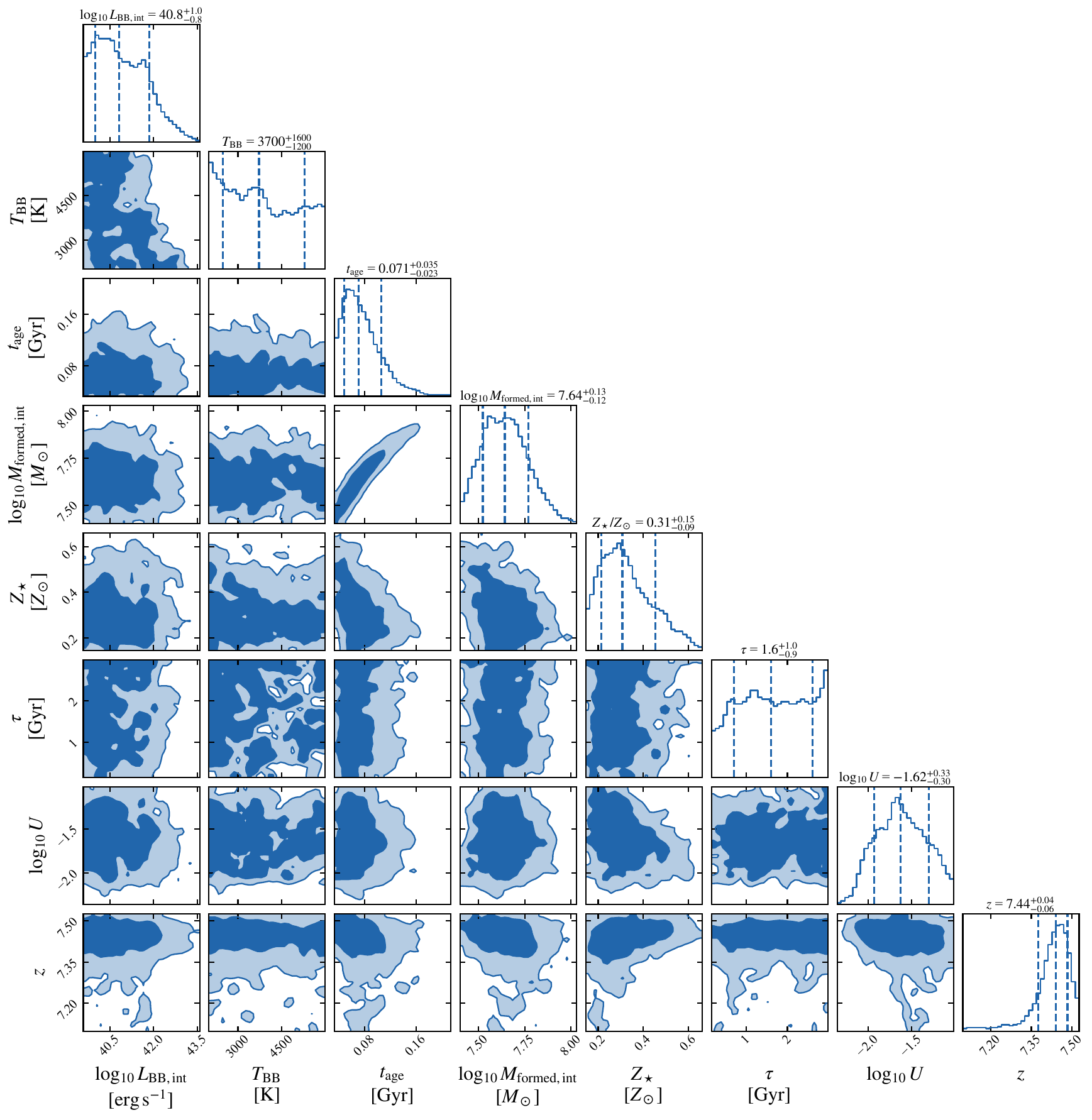}
    \caption{ SED-fitting posterior distribution for the ``unlensed blue'' $0.3''$-diameter aperture using the galaxy-plus-blackbody model shown in Figure~\ref{fig:sed-unlensed}.}
    \label{fig:corner-figure11-blue-bb}
\end{figure*}

\begin{figure*}[p]
    \centering
    \includegraphics[width=\textwidth]{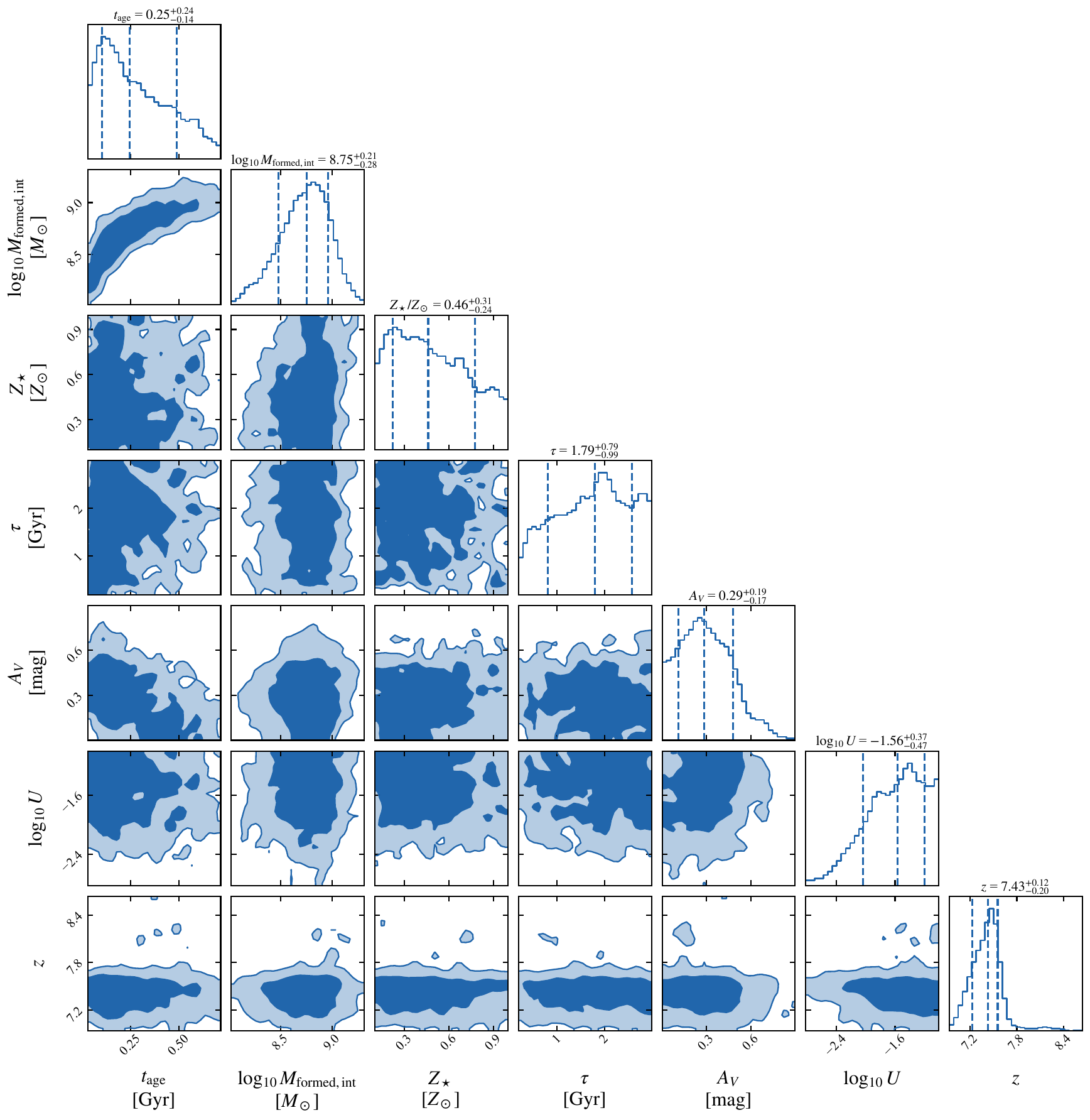}
    \caption{ SED-fitting posterior distribution for the ``unlensed total'' whole system using the galaxy-only model shown in Figure~\ref{fig:sed-unlensed}.}
    \label{fig:corner-figure11-whole-gal}
\end{figure*}

\begin{figure*}[p]
    \centering
    \includegraphics[width=\textwidth]{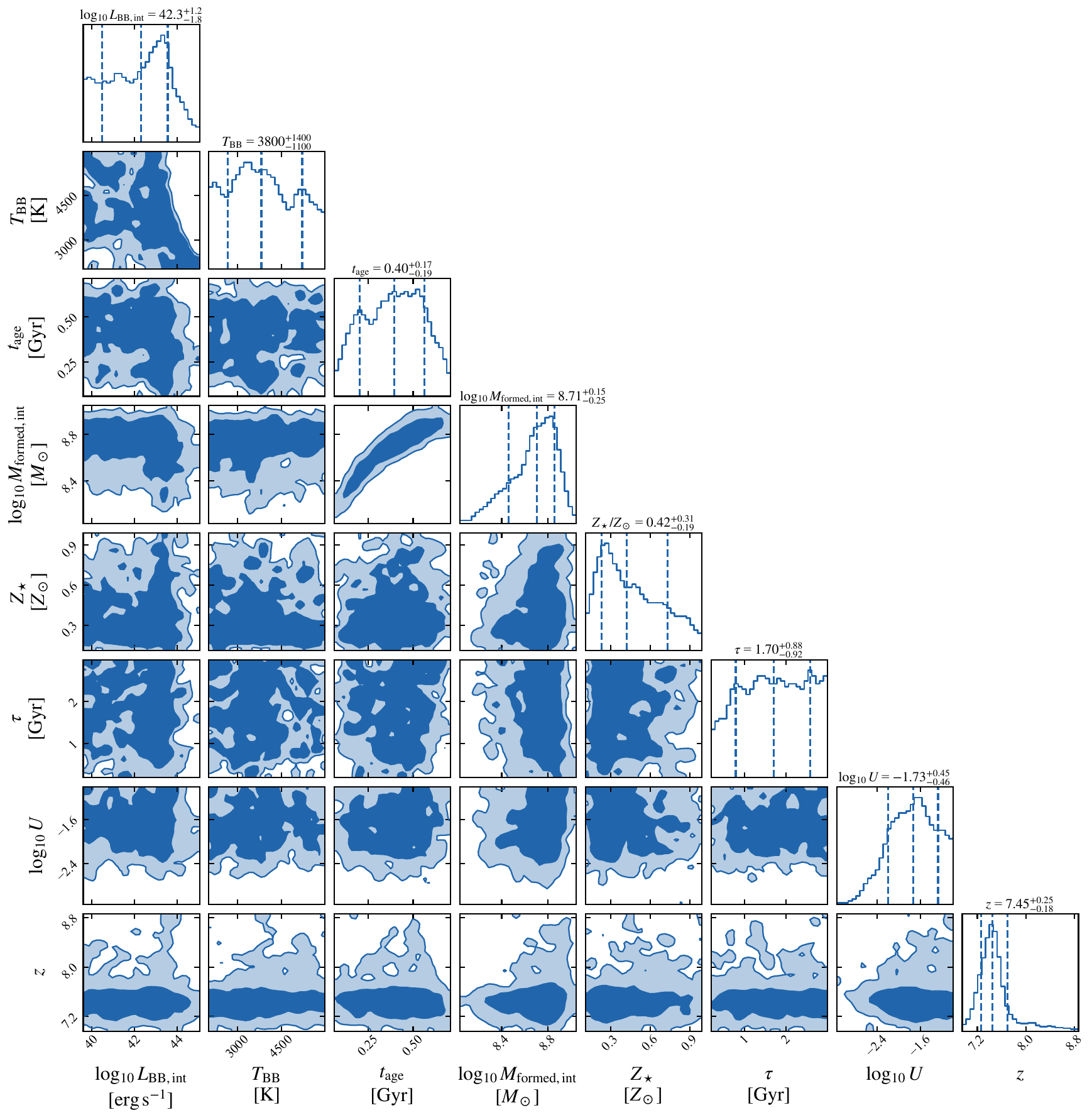}
    \caption{ SED-fitting posterior distribution for the ``unlensed total'' whole system using the galaxy-plus-blackbody model shown in Figure~\ref{fig:sed-unlensed}.}
    \label{fig:corner-figure11-whole-bb}
\end{figure*}

{
Figures~\ref{fig:corner-lrd-sw1}--\ref{fig:corner-sfg-sw6} show the marginalized posterior distributions for all 11 image-plane components used in the SED analysis. Dashed lines in the diagonal panels mark the 16th, 50th, and 84th percentiles, while the off-diagonal panels show the 68\% and 95\% credible regions. The luminosity and formed-mass normalization parameters are shown as fitted to the observed photometry, without magnification correction; the corresponding magnification-corrected values are reported in Table~\ref{tab:sed-fitting}.

Figures~\ref{fig:corner-figure8-lrd1}--\ref{fig:corner-figure11-whole-bb} show the corresponding posterior distributions for all fits displayed in Figures~\ref{fig:sed-int} and \ref{fig:sed-unlensed}. These SEDs are already intrinsic or unlensed, so no additional magnification correction is applied to their luminosity or formed-mass parameters. 
}

\clearpage

\clearpage

\clearpage

\clearpage

\clearpage

\clearpage

\clearpage

\clearpage

\clearpage

\clearpage

\clearpage

\clearpage

\clearpage

\clearpage

\clearpage

\clearpage

\clearpage

\clearpage

\clearpage

\clearpage

\begin{acknowledgments}
{We thank Sunmyon Chon, Shingo Hirano, Andrea Ferrara, Daniel P. Stark, Takumi Tanaka, Makoto Ando, Hamid Hamidani, Daisuke Toyouchi, Kazuyuki Sugimura, Fumi Egusa, Hidenobu Yajima, and Kris Boonmee} for valuable discussions.
This work is based on observations made with the NASA/ESA/CSA James Webb Space Telescope. The data were obtained from the Mikulski Archive for Space Telescopes (MAST) at the Space Telescope Science Institute (STScI), which is operated by the Association of Universities for Research in Astronomy (AURA), Inc., under NASA contract NAS 5-03127 for JWST. The JWST observations are associated with program \#6882 (VENUS) and \#1345. This paper is also based on observations made with the NASA/ESA Hubble Space Telescope. The data were obtained from MAST at STScI, which is operated by AURA under NASA contract NAS 5-26555. The HST observations are associated with program \#14096 (RELICS). The authors acknowledge the use of the Canadian Advanced Network For Astronomy Research (CANFAR) Science Platform operated by the Canadian Astronomy Data Center (CADC) and the Digital Research Alliance of Canada (DRAC) with support from the National Research Council for Canada (NRC), the Canadian Space Agency (CSA), CANARIE, and the Canadian Foundation for Innovation(CFI). 

We acknowledge the support of the Canadian Space Agency (CSA) [25JWGO4A18]. 
HY acknowledges support by KAKENHI (25KJ0832) through Japan Society for the Promotion of Science (JSPS). SF acknowledges support from the Dunlap Institute, funded through an endowment established by the David Dunlap family and the University of Toronto. MO acknowledges the supports from the World Premier International Research Center Initiative (WPI Initiative), MEXT, Japan, the joint research program of the Institute for Cosmic Ray Research (ICRR), the University of Tokyo, and KAKENHI (20H00180, 21H04467, 25H00674) through Japan Society for the Promotion of Science (JSPS). VK acknowledge support from the University of Texas at Austin Cosmic Frontier Center. MO is supported by JSPS KAKENHI Grant Numbers JP25H00662, JP22K21349.
AZ acknowledges support by the Israel Science Foundation Grant No. 864/23. MN acknowledges support from KAKENHI Grant Nos. 25KJ0828 through Japan Society for the Promotion of Science (JSPS). UH acknowledges support through grants KAKENHI (23KJ0646) through Japan Society for the Promotion of Science (JSPS). HA acknowledges support from CNES, focused on the JWST mission, and the French National Research Agency (ANR) under grant ANR-21-CE31-0838. FEB acknowledges support from ANID-Chile BASAL CATA FB210003, FONDECYT Regular 1241005, and ECOS-ANID ECOS240037. 
JMD acknowledges the support of projects PID2022-138896NB-C51 (MCIU/AEI/MINECO/FEDER, UE) Ministerio de Ciencia, Investigaci\'on y Universidades and SA101P24.
YH acknowledges support from the Japan Society for the Promotion of Science (JSPS) Grant-in-Aid for Scientific Research (24H00245) and the JSPS International Leading Research (22K21349). K.I. acknowledges support from the National Natural Science Foundation of China (12573015, W2532003), the Beijing Natural Science Foundation (IS25003), and the China Manned Space Program (CMS-CSST-2025-A09). RPN is grateful for the generous support of Neil and Jane Pappalardo through the MIT Pappalardo Fellowship. PD warmly acknowledges support from an NSERC discovery grant (RGPIN-2025-06182). RA acknowledges support of Grant PID2023-147386NB-I00 funded by MICIU/AEI/10.13039/501100011033 and by ERDF/EU. 
GEM acknowledges the Villum Fonden research grants 37440 and 13160.
MB acknowledges support from the ERC Grant FIRSTLIGHT and Slovenian national research agency ARIS through grants N1-0238 and P1-0188.
The Cosmic Dawn Center (DAWN) is funded by the Danish National Research Foundation under grant DNRF140. EV acknowledges financial support through grants INAF GO Grant 2024 ``Mapping Star Cluster Feedback in a Galaxy 450 Myr after the Big Bang'' and the European Union -- NextGenerationEU within PRIN 2022 project n.20229YBSAN - Globular clusters in cosmological simulations and lensed fields: from their birth to the present epoch.
MGO acknowledges financial support from the State Agency for Research of the Spanish MCIU through Center of Excellence Severo Ochoa award to the Instituto de Astrofísica de Andalucía CEX2021-001131-S funded by MCIN/AEI/10.13039/501100011033, and from the grant PID2022-136598NB-C32 “Estallidos8”. MGO also acknowledges the support by the project ref. AST22\_00001\_Subp\_11 funded from the EU – NextGenerationEU, PPCC Junta de Andalucía. PAAL thanks the support from CNPq, grants 310260/2025-6 and 404160/2025-5, and FAPERJ, grant E-26/200.545/2023. RAW acknowledges support from NASA JWST Interdisciplinary Scientist grants
NAG5-12460, NNX14AN10G and 80NSSC18K0200 from GSFC.

\end{acknowledgments}

\facilities{HST, JWST}

\software{
\texttt{BAGPIPES} \citep{Carnall+2018},
\texttt{BPASS} \citep{Stanway_Eldridge_2018},
\texttt{Cloudy} \citep{Gunasekera+2023},
\texttt{grizli} \citep{Brammer_2021, Brammer_2023},
\texttt{Lenstronomy} \citep{Birrer+2018, Birrer+2021},
\texttt{Lenstruction} \citep{Yang+2020},
\texttt{Photutils} \citep{Bradley+2025}
          }

\bibliography{ref}{}
\bibliographystyle{aasjournalv7}

\end{document}